\documentclass[11pt]{article}
\usepackage{jheppub}
\usepackage[utf8x]{inputenc}
\pdfoutput=1
\usepackage{graphicx}
\usepackage{subfigure}
\usepackage{amsmath}
\usepackage{bm} 
\usepackage{xspace}
\usepackage{relsize}
\usepackage{placeins}
\usepackage{multirow}
\usepackage{soul}
\usepackage{ulem}

\def\babar{\mbox{\slshape B\kern-0.1em{\smaller A}\kern-0.1em 
  B\kern-0.1em{\smaller A\kern-0.2em R}}}
\usepackage{lmodern}

\usepackage{color}
\usepackage[usenames,dvipsnames,svgnames,table]{xcolor}
\usepackage{hyperref} 
\hypersetup{linktocpage=true}
\usepackage[all]{hypcap}

\newcommand{\GeV}{{\rm GeV}}
\newcommand{\MeV}{{\rm MeV}}

\numberwithin{equation}{section}
\newcommand{\beq}{\begin{equation}}
\newcommand{\eeq}{\end{equation}}
\newcommand{\bea}{\begin{eqnarray}}
\newcommand{\eea}{\end{eqnarray}}
\newcommand{\bear}{\begin{eqnarray}}
\newcommand{\eear}{\end{eqnarray}}

\newcommand{\ba}{\begin{array}}
\newcommand{\ea}{\end{array}}

\newcommand{\BR}{{\rm BR}}

\newcommand{\pizero}{{\pi^0}}
\newcommand{\pip}{{\pi^+}}
\newcommand{\pim}{{\pi^-}}

\newcommand{\Ma}{M_a}
\newcommand{\Mpi}{M_\pi}
\newcommand{\Meta}{M_\eta}

\newcommand{\Kpi}{K_\pi}
\newcommand{\Keta}{K_\eta}

\def\KL{$K_L$ }

\usepackage{slashed}

\begin{document}

\title{\boldmath 
 KOTO vs. NA62  Dark Scalar Searches}

\author[a,b]{Stefania Gori,}
\author[c]{Gilad Perez,}
\author[d,e]{Kohsaku Tobioka}

\affiliation[a]{Santa Cruz Institute for Particle Physics, University of California, Santa Cruz, CA 95064, USA}
\affiliation[b]{Department of Physics, 1156 High St., University of California Santa Cruz, Santa Cruz, CA 95064, USA}
\affiliation[c]{Department of Particle Physics and Astrophysics, Weizmann Institute of Science, Rehovot 761001, Israel}
\affiliation[d]{Department of Physics, Florida State University, Tallahassee, FL 32306, USA}
\affiliation[e]{High Energy Accelerator Research Organization (KEK), Tsukuba 305-0801, Japan}

\emailAdd{sgori@ucsc.edu}
\emailAdd{gilad.perez@weizmann.ac.il}
\emailAdd{ktobioka@fsu.edu}

\abstract{
The two kaon factories, KOTO and NA62, are at the cutting edge of the intensity frontier, with an unprecedented numbers of long lived and charged Kaons, $\sim 10^{13}$,  being measured and analyzed.
These experiments have currently  a unique opportunity to search for dark sectors.
In this paper, we demonstrate that searches done at KOTO and NA62 are complementary, both probing uncharted territories.
We consider two qualitatively different physics cases. 
In the first, we analyze models of axion-like-particles (ALP) which couple to gluons or electroweak gauge bosons. In the second, we introduce a model based on an approximate strange flavor symmetry that leads to a strong violation of the Grossman-Nir bound.
For the first scenario, we design a new search strategy for the KOTO experiment, $K_L\to\pi^0 a\to 4\gamma$. Its expected sensitivity on the branching ratio is at the level of $10^{-9}$. This demonstrates the great potential of KOTO as a discovery machine. In addition, we revisit other bounds on ALPs from Kaon factories, highlighting the main sources of theoretical uncertainty, and collider experiments, and show new projections. 
For the second scenario, we show that the model may be compatible with the  preliminary analysis of the KOTO-data that shows a  hint for New Physics.}

\maketitle
\section{Introduction}
The Standard Model (SM) of particle physics is a successful description of Nature, especially given the discovery of the Higgs boson at the LHC~\cite{Aad:2012tfa,Chatrchyan:2012xdj}. The SM describes forms of matter which interact via the electro-magnetic, weak and strong forces. However, the SM is incomplete as it can not account for e.g. the observed baryon asymmetry of the universe, neutrino masses and mixings, and the origin of Dark Matter (DM). 
Motivated by the fine-tuning problem of the electroweak (EW) scale that conventionally requires TeV new physics (NP) which also characterizes the DM sector, tremendous efforts have been made to search for new states at the energy frontier, and yet, so far there is no conclusive sign of the beyond the SM (BSM) physics.
On the other hand, a NP sign could appear as a light weakly coupled state, for instance associated with a pseudo Nambu Goldstone boson (pNGB) field, and the representative example is an axion or axion-like-particle (ALP)\footnote{A terminology of axion-like-particles is not well-defined. Here we use it as a light CP-odd particle with couplings to gauge bosons and with a mass not uniquely determined by its decay constant. }. 
The mass scale of the pNGB can be substantially lighter than the GeV scale, and its interaction strength with SM particles can be suppressed by a higher symmetry-breaking scale. This type of particle can be tested at high-intensity experiments, such as in rare meson decay measurements, at B-factories, beam-damp experiments, and neutrino experiments.  

Among the high intensity experiments, the Kaon factories,  KOTO and NA62 experiments, are unique since they aim to measure Kaon decays with a branching ratio as small as $\sim 10^{-11}$, collecting an extraordinary large number of Kaon decays, $\sim 10^{13}$. More specifically, the KOTO experiment aims to detect for the first time the SM decay, $K_L\to \pi^0\nu\bar\nu$, while the NA62 is searching for the charged counterpart, $K^+\to \pi^+\nu\bar\nu$. The SM prediction for the branching ratios is tiny, $\BR(K_L\to \pi^0\nu\bar\nu)=(3.00\pm 0.30)\times 10^{-11}$ and $\BR(K^+\to \pi^+\nu\bar\nu)=(9.11\pm 0.72)\times 10^{-11}$ \cite{Brod:2010hi,Buras:2015qea}. Given the very small branching ratios, these decays are extremely sensitive to NP effects. Under some fairly general assumptions, discussed below, the charged and neutral decay channels are tightly connected leading to the Grossman-Nir (GN) bound \cite{Grossman:1997sk}, $\BR(K_L\to \pi^0   \nu\bar\nu) \lesssim 4.3~ \BR(K^+\to \pi^+  \nu\bar\nu)$, which may hold even if the final state is modified but the topology remains similar as is further discussed below (see the recent discussion in~\cite{Kitahara:2019lws,He:2020jzn,Jho:2020jsa,Dev:2019hho,Fabbrichesi:2019bmo,Mandal:2019gff}. For earlier discussions on the violation of the GN bound see~\cite{Buras:2004uu,Fuyuto:2014cya,Hou:2015ckg,Hou:2016den,Grossman:2003rw}). 

In this paper, we demonstrate that both Kaon factories have great opportunities as discovery machines of light new particle. In Sec.~\ref{sec:LightScalar}, we introduce two qualitatively different physics cases that show that NP searches done via charged Kaon decays at NA62 and via neutral Kaon decays at KOTO are complimentary, as opposed to be strongly linked with each other. 
This is in contrast to what one would naively expect to be the case due to the GN bound.

First, we consider an ALP ($a$) with coupling to gluons or $W$ bosons as a representative candidate of pNGB. In this context, we propose a novel search for the KOTO experiment (Sec. \ref{sec:KOTOsearch}). Specifically, KOTO can search for $K_L\to \pi^0  a$ where the ALP decays to di-photon. {This search will be complementary to the $K_L\to \pi^0  a$ with an invisible ALP search that is already performed by the collaboration. These two channels together will probe experimentally unexplored parameter space of the ALP coupled to SU(2) gauge bosons (Sec. \ref{Sec:SU(2)}) or to gluons (Sec. \ref{subsec:GG}), in the mass range from 10~MeV to 350~MeV. NA62 will also probe parameter space through the corresponding $K^+\to \pi^+  a$ decay that we analyze.}

The second scenario we analyze in this paper is a theory with an approximate strange flavor symmetry, with an additional light, flavon-like, complex scalar field, $\phi$. We discuss its phenomenology in Sec.~\ref{sec:GNbreaking}.  
The flavor preserving coupling allow for a SM singlet final state, consisting of the real ($\sigma$) and imaginary ($\chi$) parts of $\phi$, to
be accessible only to $K_L$ and not to its charged isospin-partner. Therefore, breaking the GN relation. KOTO is particularly sensitive to such a scenario, once we allow the $\chi$ to decay to two photons. Furthermore the expected signal can be made compatible 
with the preliminary analysis of the KOTO-data that shows a hint for NP \cite{KOTOslides} (though more investigation of the collaboration is needed). Other explanations of this anomaly can be found in \cite{Kitahara:2019lws, Mandal:2019gff, Calibbi:2019lvs, Li:2019fhz, He:2020jzn, Fabbrichesi:2019bmo, Egana-Ugrinovic:2019wzj, Dev:2019hho, Liu:2020qgx, Banerjee:2020kww, Jho:2020jsa}.

\section{Light Scalars at Kaon Factories} \label{sec:LightScalar}
Here, we describe the two new physics scenarios where the Kaon factories can play a major role probing the parameter space.

\subsection{Massive axions, ALPs and pNGBs}
The Goldstone theorem provides one of the most compelling motivation for the presence of light scalars as their masses are protected by a shift symmetry.
The simplest manifestation of the Goldstone theorem is the case of a spontaneously broken U(1) symmetry that leads to the presence of a light ALP.
Such a state can be motivated by a solution of the hierarchy problem~\cite{Graham:2015cka}, the strong CP problem~\cite{Peccei:1977hh,Wilczek:1977pj,Weinberg:1977ma}, the flavor puzzle~\cite{Froggatt:1978nt},  and combinations of these with DM physics~\cite{Wilczek:1982rv,Abbott:1982af,Dine:1982ah,Preskill:1982cy,Calibbi:2016hwq,Ema:2016ops,Banerjee:2018xmn}.
For concreteness, to motivate our scenario we focus on the QCD axion case, however, the essence of our reasonings below holds for a broader class of ALP models.
The typical breaking scale of the Peccei-Quinn (PQ) symmetry \cite{Peccei:1977hh}, $F_a$, considered in literature is rather high. The standard axion window is 
$10^9\lesssim F_a \lesssim 10^{12}~\rm GeV$.%
The upper bound is due to the over-production of axion as dark matter, and the lower bound comes from astrophysical observations\cite{Kim:2008hd}.

However, there is a theoretical concern about the quality of the PQ symmetry with a high decay constant~\cite{Georgi:1981pu,Lazarides:1985bj,Kamionkowski:1992mf,Holman:1992us,Barr:1992qq}. Any global symmetry is believed to be broken by the UV physics: the quantum gravity does not respect global symmetries; or any global symmetry can be an accidental symmetry of the UV physics.    
In the effective field theory, this conjecture implies that higher dimensional operators suppressed by the UV physics scale, ${\Phi |\Phi|^{D-1}}/{\Lambda^{D-4}_{UV}}$, can explicitly break the PQ symmetry 
where $\Phi$ is a field which carries a non-zero PQ charge and has a VEV of $F_a$. These operators  ruin the PQ mechanism because this operator shifts the minimum of the axion potential away from $\bar\theta=0$, 
\begin{align}
V(a)&= m_a^2F_a^2 \left\{1-\cos \left(\frac{a}{F_a}\right) \right\}+\frac{F_a^2}{\Lambda_{UV}^{D-4}}\cos\left(\frac{a}{F_a}+\Delta\right)
\label{eq:quality1}
\\
\to\delta\bar\theta&=\frac{\delta a_{min}}{F_a} \sim \frac{F_a^{D-2}}{m_a^2 \Lambda_{UV}^{D-4}}\,,
\label{eq:quality2}
\end{align}
where $\Delta$ is a non-aligned CP phase that is generically expected to be of order one.  
Even though the deviation is suppressed by a high scale $\Lambda_{UV}\leq M_{\rm pl}$, the effect in the $\bar\theta$ can be  significant because of two factors: (1) the original axion potential is not very steep, $m_a^2 F_a^2\approx m_\pi^2 F_\pi^2$; (2) the precision of the neutron EDM measurement is accurate, $\delta\bar\theta\lesssim 10^{-10}$. 
Therefore, operators up to $D\simeq 10$  need to be absent to maintain the PQ mechanism. 
This situation is unsatisfactory from the low energy point of view. Some mechanism should maintain the quality of the global PQ symmetry to be extremely good to solve the strong CP problem. 
This problem is not unique to the QCD axion but is also common to other solution to the QCD CP problem~\cite{Dine:2015jga} and other mechanisms that strongly rely on precise global symmetries~\cite{Higaki:2016yqk,Davidi:2017gir,Cox:2019rro}.

\subsection*{Heavy Axion as a Consequence of the  Quality Problem}
To construct theories that are protected against Planck suppressed operators of $D\geq5$, the favored decay constant is necessarily low. 
Assuming the standard relation of axion mass and decay constant, $m_a\approx m_\pi F_\pi/F_a$,  and requiring a small deviation, $\delta\bar\theta<10^{-10}$, one can obtain the  bound  on the effective decay constant and the mass, 
\begin{align}
F_a \lesssim10~{\rm GeV} \quad {\rm and}\quad m_a\gtrsim 1~{\rm MeV}\,.
\end{align}
This parameter space is similar to the original Weinberg-Wilczek axion \cite{Wilczek:1977pj,Weinberg:1977ma}. This motives us to search for axions with a mass at and above the MeV scale.  
 
The low decay constant  along the standard QCD  axion relation has been excluded by astrophysical observations and beam-dump experiments. However, the bounds do not apply if there is an additional contribution to the axion mass.  
Many phenomenological studies for ALPs  show that parameter space with very low decay constant are poorly constrained if the mass is heavier than $\sim 50$~MeV (see recent works, for example, Refs.~\cite{Bauer:2017ris,Mariotti:2017vtv, Dolan:2017osp, CidVidal:2018blh, Aloni:2018vki, Aloni:2019ruo, Hook:2019qoh, Gavela:2019cmq}). 
Indeed, there are models of heavy axions without the standard axion relation, where the strong CP problem can be addressed~\cite{Rubakov:1997vp,Fukuda:2015ana,Hook:2019qoh,Gaillard:2018xgk,Gherghetta:2016fhp,Agrawal:2017ksf,Higaki:2016yqk}. 
As an example of non-minimal heavy axion, we consider a scenario with a mirror strong sector. The mirror sector shares a strong CP phase and quark phases with the SM ensured by a $Z_2$ symmetry, and a single axion relaxes the two CP phases of the SM and mirror sectors. 
With soft breaking of the $Z_2$ symmetry, a higher confinement scale in the mirror sector is achieved, which is an extra source of the axion mass. 

We now revisit the quality problem and the bound on the axion mass based on the scenario with the mirror strong sector. First, the higher dimensional operator should be sufficiently suppressed as in Eq.~ \eqref{eq:quality2}, 
\begin{align}
F_a\lesssim \left(m_a^2 \Lambda_{UV}^{D-4} \delta\bar\theta \right)^{\frac{1}{D-2}}. 
\label{eq:fabound1}
\end{align}
The axion mass is dominated by the contribution from the mirror sector because its confinement scale $\Lambda'$ is much higher than the SM one, $\Lambda_{\rm QCD}$,  
\begin{align}
m_a^2 \simeq \frac{m_q \Lambda'^3}{f^2_a} +{\cal O}\left(\frac{m_\pi^2 F_\pi^2}{f^2_a}\right)
\end{align}
{where $m_q$ are SM quark masses.}  Generally, we expect a hierarchy $\Lambda' \ll F_a$ because $\Lambda'$ is generated by dimensional transmutation, but  the confinement scale can be up to $\Lambda'=F_a$, and consequently $m_a^2< m_q F_a  $. 
Combining this with  Eq.~\eqref{eq:fabound1}, we get
\begin{align}
m_a <\left( m_q  \Lambda_{UV}^{\frac{D-4}{D-2}}\delta\bar\theta^{\frac{1}{D-2}}  
\right)^{\frac{D-2}{2(D-3)}}. 
\end{align}
Specifically, the bounds for $D=5$ case are
\begin{align}
& 1~{\rm MeV}\lesssim  m_a\lesssim
5~{\rm GeV}
\left(\frac{m_q}{10~\rm MeV}\right)^{\frac{3}{4}}
\left(\frac{\Lambda_{UV}}{M_{\rm pl}}\right)^{\frac{1}{4}}
\left(\frac{\delta\bar\theta}{10^{-10}}\right)^{\frac{1}{4}}, 
\\
&F_a\lesssim200~{\rm GeV}\left(\frac{m_a}{100~\rm MeV}\right)^{\frac{2}{3}}
\left(\frac{\Lambda_{UV}}{M_{\rm pl}}\right)^{\frac{1}{3}}
\left(\frac{\delta\bar\theta}{10^{-10}}\right)^{\frac{1}{3}}. 
\end{align}
The above parameter space is only weakly covered by the current experimental probes. 
Since part of this mass range is within the range of Kaon experiments (particularly the lower mass range), it is very important to develop a search program to discover heavy ALPs at Kaon factories. 
For the phenomenological study of heavy axions, we consider two simplified models: the first involves a ALP coupled to the electroweak sector of the SM, and the second a ALP coupled to gluons (for more information see Sec.~\ref{sec:simplified}). 

\subsection{The generalized GN bound and how to avoid it}\label{Sec.2.2}
 Under fairly general assumptions, 
the $K_L \to \pi^0 \nu\bar\nu$ rate can be strongly constrained by the $K^+ \to \pi^0 \nu\bar\nu$ rate via the Grossman-Nir~(GN) bound~\cite{Grossman:1997sk}:
\begin{align}
	\label{eq:GNbound}
	\BR(K_L\to\pi^0\nu\bar\nu) 
	\ \leq \
	4.3 \, \BR (K^+\to\pi^+\nu\bar\nu)\,.
\end{align}
The numerical factor comes from the difference in the total decay widths of $K_L$ and  $K^+$, isospin breaking effects, and QED radiative corrections~\cite{Mescia:2007kn, Buras:2015qea}.
The GN bound only relies on the following assumptions~\cite{Grossman:1997sk}: 
First, the isospin symmetry, which relates the decay amplitudes of $K^\pm$ to the ones of $K^0$ and $\bar K^0$.  
Second, the ratio of the $K$ and $\bar K^0$ decay amplitudes to the corresponding sum of final states is close to unity, where if the final state is CP eigenstate it means no CPV in the decay. For the $\pi\nu\bar\nu$ final state, within the SM, it is expected to be an excellent approximation.
The above assumptions are not easy to be violated even when going beyond the SM. 

Inspired by~\cite{talkbyMP}, we shall construct a model based on an approximate global flavor symmetry, that avoids the GN bound via exploiting strong isospin breaking (see~\cite{Buras:2004uu,Fuyuto:2014cya,Hou:2015ckg,Hou:2016den,Grossman:2003rw} for relevant discussions).
To realize the idea, we add a light complex scalar, $\phi$, which carries a half strange (or second generation doublet) flavor charge. 
This implies that we expect the following operator to be allowed by the symmetry and present in the effective theory, in the down quark mass basis, 
\beq
y_1 H\bar Q_1 s \phi^2/\Lambda^2 \ \ {\rm and/or} \ \ y_2 H\bar Q_2 d \phi^2/\Lambda^2+h.c.\,,\label{EqGN}
\eeq
where the first (second) operator corresponds to $\phi^2$ carries a unit $\bar s$ ($Q_2$) flavor charge, and we assume $\langle \phi\rangle=0$.
In the broken electroweak phase, this effective Lagrangian leads to an effective operator  $ y_{1,2}\bar s d \phi^2+h.c.$ that induces the $K_L\to \sigma \chi$ decay, with $\sigma = {\rm Re}(\phi){/{\sqrt{2}}}$ and $\chi={\rm Im} (\phi){/{\sqrt{2}}}$ (here, for simplicity, we assume an approximate CP conservation in the decay). 
Using NDA, from Eq.~\eqref{EqGN} we expect 
\beq
\Gamma(K_L\to \chi\sigma) \sim M_K \left | y_{1,2} v \over\Lambda^2\right|^2 \times F_{\pi}^2 \,.
\eeq
However, due to conservation of charge there is no analogous 2-body decay of the charged Kaon unless additional charge pions are added to the final state.
This implies that the charged Kaon decay is suppressed, by two-vs-three-body (and possibly kinematical) phase space factors which implies a strong violation of the effective new physics GN bound. 
As discussed in Sec. \ref{sec:GNbreaking}, we find that the NP charged Kaon decays are suppressed by at least two orders of magnitude relative to the $K_L$ one. Thus, in such a scenario, it is possible that while, at present, the KOTO detector is sensitive to a NP signal, the NA62 one is not.

The model, as presented above, has an exact $\phi$-parity symmetry which renders the $\phi$ state stable. To achieve a visible signal at Kaon experiments, we add a CP conserving coupling, 
\begin{equation}\label{eq:chiDecay}
\mathcal L_{\chi}\supset\frac{\chi}{\Lambda_\chi} F_{\mu\nu}\tilde F^{\mu\nu}\,,
\end{equation}
that is responsible to the decay of $\chi$ into two photons. 
Up to small symmetry breaking effects, to be discussed below in Sec. \ref{sec:GNbreaking}, $\sigma$ would be stable and hence the final state of the $K_L\to \sigma \chi(\gamma\gamma)$ is similar to the $K_L\to\pi^0\nu\bar\nu$, which KOTO is searching for. 

\section{The KOTO experiment}
\subsection{Overview}
KOTO is an experiment searching for the rare neutral Kaon decay, $K_L \to \pi^0 \nu\bar\nu$, whose branching ratio is expected to be $(3.0\pm0.3) \times 10^{-11}$ \cite{Brod:2010hi,Buras:2015qea}. In the past, the E391a experiment, at KEK, set the most stringent limit on the branching ratio at $2.6\times 10^{-8}$ \cite{Ahn:2009gb}. The first KOTO analysis based on data collected in 2015 was able to set a bound at BR$(K_L \to \pi^0 \nu\bar\nu)_{\rm{KOTO}}<3\times 10^{-9}$ \cite{Ahn:2018mvc}. This is relatively close to the bound obtained from the charged decay, $K^+\to\pi^+\nu\bar\nu$, using the Grossman-Nir bound: BR$(K_L \to \pi^0 \nu\bar\nu)_{\rm{GN}}<1.46\times 10^{-9}$.

\begin{figure*}[h!]
 \begin{center}\label{KOTOExp} 
  \includegraphics[width=0.95\linewidth]{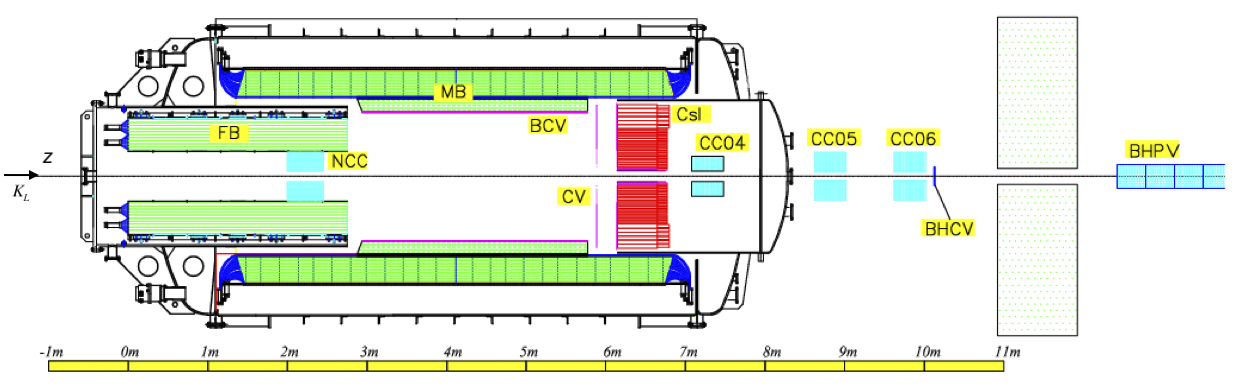}
 \end{center}
 \caption{Layout of the KOTO detector, taken from~\cite{Hewett:2012ns}. The Kaon beam enters from the left, as indicated by the arrow. 
 Schematic drawing of the detector. The components of the detector include collar counters (CCxx), Neutron Collar Counter (NCC), Front Barrel (FB), Main Barrel (MB), charged-particle vetos (BCV and CV), CsI crystals (CSI), Beam Halo Charged Veto (BHCV) and Photon Veto (BHPV).
 For more information see~\cite{Masuda:2015eta}.
 }
\end{figure*}

KOTO is a fixed target experiment that utilizes a 30~GeV proton high intensity beam extracted from the J-PARC main ring accelerator. The produced Kaons are purified by a 20m-long beam line and enter in the detector of Fig.~\ref{KOTOExp}, as indicated by the arrow, where the beam axis is denoted as the $Z$ direction.
The flux of Kaons was measured by an engineering run in 2015 at  $Z\sim -1.5$m~\cite{Masuda:2015eta}.
The actual detector consists of a CsI calorimeter (Ecal) at the front target and various veto detectors for charged particles and photons.

\begin{figure*}[h!]
 \begin{center}
  \includegraphics[width=0.5\linewidth]{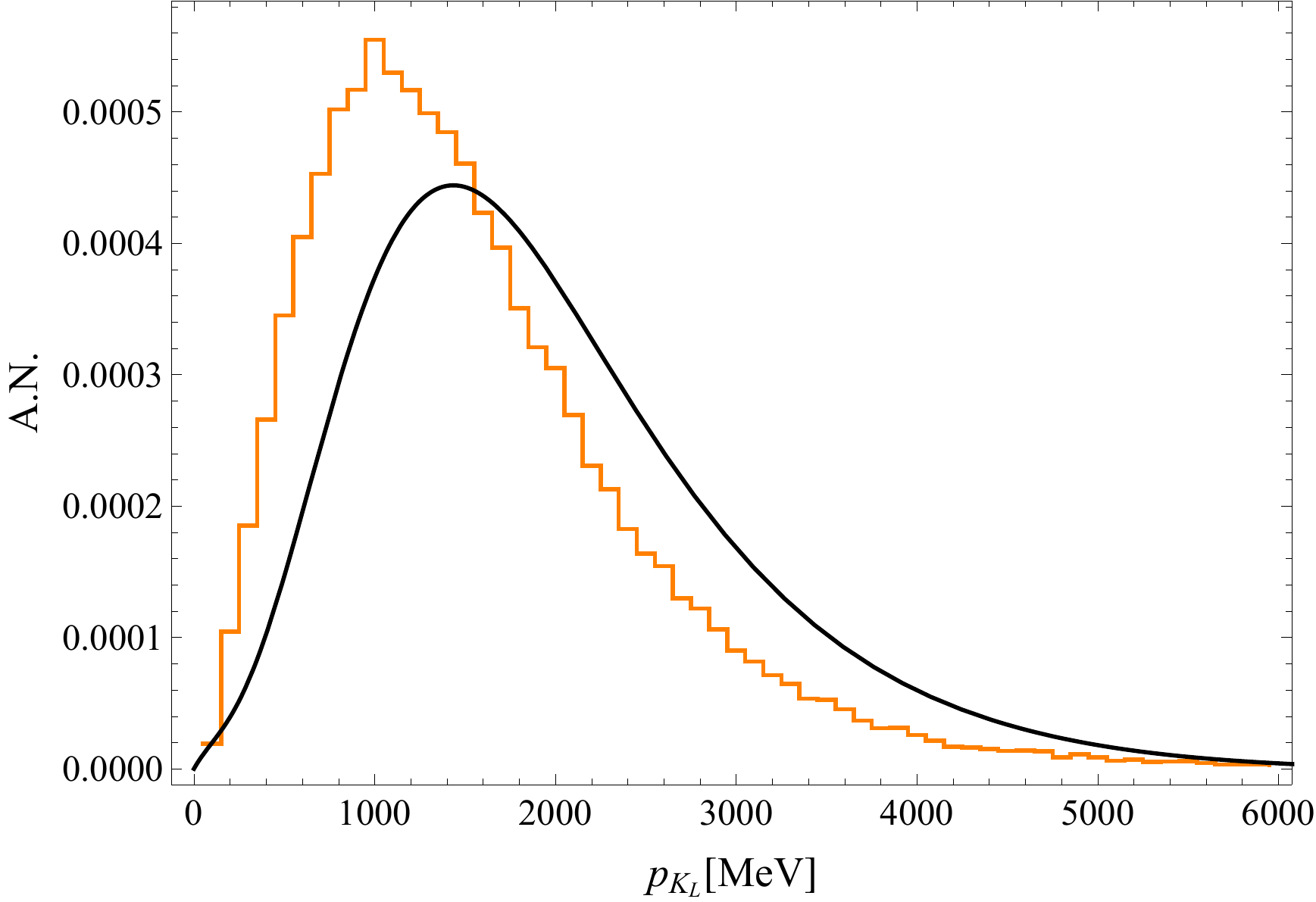}
 \end{center}
 \caption{
 Left: The solid black line represents the measured $K_L$ momentum distribution in a special run at the beam exit (Fig.~6 of \cite{Masuda:2015eta}). The histogram shows the momentum distribution of  $K_L$ decaying within the detector (the normalization is arbitrary). 
 }\label{KLmom}
\end{figure*}

The measured momentum distribution of the incoming \KL  flux is shown in black in Fig.~\ref{KLmom} and it peaks at around 1.5 GeV. Then the Kaons decay in the decay volume at $2{\rm{\,m\,}}<Z< 6.148$\,m to produce pions or neutrinos, and the momentum distribution of the decayed \KL is shifted towards lower values as shown by the orange histogram in Fig.~\ref{KLmom}.  The neutral pions are reconstructed through the identification of photons that hit the CsI calorimeter with $E_\gamma>2$\,MeV. 

The output from the detector is the position of the photon energy deposition on the ECAL (on the plane perpendicular to the beam direction), and the timing of the hits. What is known is the energy of photons rather than their four-momenta because the decay vertex of the Kaon (effectively same as the pion), $Z_{\rm vtx}$, is unknown and  the ECAL can measure only the photon energy. Furthermore, the final states of interest include no charged particles, which could provide directional information.   
In order to reconstruct the decay vertex, the standard technique  is to impose at least one additional assumption regarding the invariant mass of the parent particle or intermediate particles \cite{Abouzaid:2008xm, Masuda:2015eta} (see appendix \ref{sec:SMrecon} for a brief review). This procedure still has multi-fold ambiguities, but the correct vertex can be picked, at least based on statistical merit, by requiring that the reconstructed event describes the physical process. 
This challenge holds for SM processes such as $K_L \to \pi^0(\gamma\gamma)\nu\bar \nu$, $K_L \to \pi^0 \pi^0 \to 4\gamma$, as well as possible new physics processes, to be discussed below.

\subsection{Flux, signals, and future plans}
In Tab.~\ref{flux}, we summarize the most important numbers that characterize the intensity of the KOTO experiment. We compare the amount of data collected in 2015, to the one collected in 2016-2018. We then report the amount of data that is aimed to be collected in the coming years to reach the measurement of several SM events for $K_L \to \pi^0 \nu\bar\nu$.

The $K_L$ flux at the beam exist is usually reported. We refer to it as $N^0_{K_L}$, which is calculated by protons on target (POT)~\cite{Masuda:2015eta}, 
\begin{align}
&2\times 10^{14}\ {\rm POT}=(4.188\pm 0.017) \times 10^7 \ {\rm  Kaons\ at\ beam\ exit}.  \label{POTtoKL}
\end{align}
The reduction factor of $\sim 10^{-7}$ in Eq.~\eqref{POTtoKL} can be understood in this way: the number of  produced Kaons at the target is $10^{-1}$--$10^{-2}$ per proton (for $\cal{O}$(100) GeV); the fraction of Kaons that survive until the beam exit is $\exp[-20{\rm m}/c\tau_{K_L}]\sim 25\%$; $K_L$ need to travel through the beam hole 20m away from the production point, and the corresponding effective angle is small, $\Delta \Omega/\Omega\sim (14.8{\,\rm cm}\times 14.8{\,\rm cm})/(20{\, \rm m} \times 20 {\, \rm m})\sim 10^{-4}$. 

Since most of the entering Kaons do not decay, we need to translate $N^0_{K_L}$ to the fraction of decays inside the relevant detector volume. The probability for the $K_L$ to decay
inside the entire detector region, $2\,{\rm m}<Z<6.148$\,m, calculated at truth level is 7.9\%. %
The fraction of $K_L\to\pi^0 \nu\bar\nu$ relevant to the actual analysis, within the region of $3\,{\rm m}<Z<4.7$\,m, using our reconstruction-level simulation was found to be 3.2\% (consistent with~\cite{CortinaGil:2018fkc}), while the one associated with multiple pion final state analysis, which corresponds to the region of $2{\,\rm m}<Z<5.4\,$m was found to be 6.5\%, at the truth level (consistent with~\cite{Masuda:2015eta}).

\begin{table}[t]
\begin{center}
  \begin{tabular}{ |c ||r| r | r | }
  \hline
    & 2015 \cite{Ahn:2018mvc} & 2016-2018 \cite{KOTOslides} & Future \cite{KOTOslidesNanjo}\\
  \hline
  POT & $2.2\times 10^{19}$ & $3.4\times 10^{19}$& $\sim 10^{21}$
  \\ \hline
$K_L$'s entering the detector,  $ N^0_{K_L}$ & $4.62\times 10^{12}$ & $7.1\times 10^{12}$ &$\sim 2.1	\times 10^{14}$
  \\ \hline
$K_L$'s decaying in detector, $2{\rm\,m}\lesssim Z\lesssim 6.148\,$m  & $3.68\times 10^{11}$& $5.7\times 10^{11}$  &$\sim 1.6\times 10^{13}$
  \\\hline
  $K_L\to \pi^n$  within $2{\rm\,m}<Z<5.4{\rm\,m}$  & $3.02\times 10^{11}$ & $4.7\times 10^{11}$  &$\sim 1.4\times 10^{13}$ 
  \\\hline\hline
   $K_L\to \pi\nu\bar \nu$  within $3{\rm\,m} <  Z < 4.7{\rm\,m}$ & $1.48\times 10^{11}$ & $2.3\times 10^{11}$  &$\sim 6.6\times 10^{12}$
  \\\hline
  S.E.S of $K_L\to \pi \nu\bar\nu$ & $1.3\times 10^{-9}$ & ${6.9\times 10^{-10}}$ &$\sim 2.9\times 10^{-11}$
 
  \\\hline
   \end{tabular} 
  \end{center} 
  \caption{Summary of expected number of Kaons in different regions of the experiment. The number of protons on target (POT), the number of Kaons entering the detector, $N^0_{K_L}$, and single-event-sensitivity (S.E.S)  are given in the literatures, while number of Kaons decaying in the various regions is calculated with the measured $K_L$ momentum distribution. The S.E.S. does not scale by just statistics from the 2015 analysis to the 2016-2018 analysis because of an improved acceptance from $1.7\times 10^{-4}$ to $2.0\times 10^{-4}$.   For more details see text.
   }\label{flux}
\end{table}

For the following discussion, it is useful to examine the search for the $K_L\to \pi^0 \nu\bar\nu$ decay as done by the KOTO experiment in more detail.  The relation between the flux, acceptance and the S.E.S is given by 
\begin{align}
N^0_{K_L}&=1/(A_{\pi^0\nu\bar\nu}\times {\rm S.E.S.}_{\pi^0\nu\bar\nu})\,. 
\end{align}
The event selections are  given in~\cite{Ahn:2018mvc}, and most of them are included in our analysis, except the veto and shower-shape cuts. To keep these cuts into account, we choose uniform efficiencies: for the veto cut we use 0.17, and for the shower-shape cut we use 0.52~\cite{Ahn:2018mvc}. The signal region after these cuts is defined by the transverse momentum of the  reconstructed $\pi^0(\gamma\gamma)$ and its decay vertex, $Z_{\rm vtx}$, that is, $p_T^{\rm min}(Z_{\rm vtx})< p_T^{\pi^0}<250~\MeV$ and $3\,{\rm m}<Z_{\rm vtx}<4.7{\rm \,m}$\footnote{Note that in the recent KOTO analysis~\cite{KOTOslides}, a slightly different cut was employed, $3.2\,{\rm m}<Z_{\rm vtx}<5{\rm\,m}$, which we shall also adapt when comparing with this data sample.},
where $p_T^{\rm min}(Z_{\rm vtx})=130~\MeV+\max [0, 20~\MeV\left(\frac{Z_{\rm vtx}-4\rm~m}{0.7\rm\,m} \right)]$. 
{Note that the vertex reconstruction in the $K_L\to \pi^0 \nu\bar\nu$ search occasionally has two-fold physical solutions for $Z_{\rm vtx}$, so we choose the one further from the ECAL. We have checked that this is the physical one in most cases. Even if we discards the events with the two-fold solutions, the acceptance changes by only $\mathcal O(1\%)$. 

In Sec. \ref{sec:GNbreaking}, we will discuss how the requirement to have photons in the signal region affects the acceptance of a NP model that leads to a $K_L\to\sigma\chi,\chi\to\gamma\gamma$ decay.

\section{New ALP searches at KOTO}\label{sec:KOTOsearch}

The KOTO experiment can look for heavy axions or ALPs produced from $K_L$ decays. Particularly, as we will argue, searches could be designed to identify the decay topology $K_L \to \pi^0 a,~a\to\gamma\gamma$, that we will analyze in detail in this section.

\subsection{Reconstruction of the four photon signature}\label{sec:recon}
The signal of our interest is four photon final state from $K_L \to \pi^0 (\gamma\gamma) a(\gamma\gamma)$. As already mentioned, in order to reconstruct the decay vertex, the standard technique  is to impose at least one assumption on the invariant mass of the parent particle or intermediate particles. This procedure still has multi-fold ambiguities, but the correct vertex can be picked within an error of a few percent by consistency checks. However, note that this technique works only for the anticipated decay topologies. 
\subsubsection*{Axion reconstruction}
Extending the standard technique, we propose a new algorithm to reconstruct the signal process $K_L \to \pi^0 a 
\to 4\gamma$ without knowing the axion mass, as follows.
\begin{enumerate}
\item 
Divide the four photons into two pairs which can be associated to mother particles, say the pair  $A$ consists of $\gamma_1$ and $\gamma_2$ and the other pair, $B$,  consists of $\gamma_3$ and $\gamma_4$. There are six such combinations of photon-pairs. 
\item
Obtain a candidate vertex, $Z_{\rm vtx}$,  assuming that the mother of the photon-pair $B$ is a neutral pion. This assumption holds for the signal as well as physical background processes $K_L\to \pi^0\pi^0, \pi^0\gamma\gamma$. Consequently, the invariant mass of the pair $B$ can be written as, 
\begin{align}
m_{\gamma_3\gamma_4}^2(Z_{\rm vtx})\equiv m_{\pi^0}^2
\end{align}
There are up to two solutions for $Z_{\rm vtx}$. One solution is often unphysical, since the reconstructed vertex is outside from the decay volume, or it is an imaginary number, and thus discarded. 
\item
Repeat the above steps for all the possible parings, which lead to twelve-fold ambiguities in a four-photon event. 
\item
If the pairing and the reconstructed vertex are found to be physically consistent, the four photon invariant mass, $m_{4\gamma}$, is also required to be peaked around the $K_L$ mass, and the
  combination that minimizes the resulting $|m_{K_L}-m_{4\gamma}|$ is selected. 
\end{enumerate}
This reconstruction algorithm works quite well, allowing us to select the correct pair in the signal process. The fraction of the (preselected candidate) events that are correctly selected  is  90\%  (70\%  for $m_a\simeq m_{\pi^0}$). The di-photon invariant mass of one pair is expected to have a peak at around the ALP mass, while the background events {$K_L\to\pi^0\pi^0$} have a peak in the same variable around the neutral pion mass.

\subsection{Simulation}
We develop a MC simulation based on our reconstruction algorithm, and estimate the acceptance of the signal and the {$K_L\to\pi^0\pi^0,~K_L\to\pi^0\gamma\gamma$} backgrounds. We start from the known $K_L$ flux, and then let $K_L$ decay to $\pi^0 \pi^0, \pi^0 \gamma\gamma, \pi^0 a$, and subsequently decay the $\pi^0$ and $a$ to $\gamma\gamma$. For the photon energy measurements, we include the dominant smearing effects due to the ECAL.

\subsubsection*{\boldmath $K_L$ momentum and vertex reconstruction}
Our simulation aims at obtaining the two-dimensional distribution of the Kaons in terms of the reconstructed momentum and the reconstructed decay point, $Z_{\rm vtx}$. 
We generate the Kaon momentum according to its measured distribution at the beam exit $Z\simeq-1.5\,$m, shown in black in Fig.~\ref{KLmom}. The Kaons then decay according to their  lifetime of approximately 0.51\,ns.  We assume that the Kaons are fully aligned with the beam axis.
We collect  the decayed $K_L$ within the decay volume of the detector, $2{\rm\, m}<Z_{K_L}<6.1\,$m where $Z_{K_L}$ is the actual point of $K_L$ decay, and the decay probability is about 7.9\%\footnote{Backgrounds  from upstream decays are not included in our simulation.}. As a cross-check, we calculate the decay probability in the fiducial volume of $K_L\to \pi\nu\bar\nu$ analysis, that is, 3.2\%, and it is consistent with the reported probability in~\cite{CortinaGil:2018fkc}.

\subsubsection*{Decays to four photons}
Based on the distribution of decaying $K_L$ within the detector ($2\,{\rm m} \!\lesssim \! Z_{K_L}\!\lesssim \!6.1\,$m), we generate events for three decay processes,  $\pi^0\pi^0, \pi^0\gamma\gamma$, and $\pi^0a$\,. The MC sample size is $5\times 10^4$ events for the $K_L\to \pi^0 a\to4\gamma$ signal for each mass bin ($m_a=1, 10,20, ..., 360$\,MeV), and $2\times 10^5$ ($1.5\times 10^6$) events for the $K_L\to \pi^0 \gamma\gamma\to4\gamma$ ($K_L\to \pi^0 \pi^0\to4\gamma$) background. 
We treat the two-body decay processes as spherically symmetric, which is a good approximation. The three body decay $K_L\to \pi^0 \gamma\gamma$ has a non-trival Dalitz-plot distribution, thus, we take into account the shape based on the matrix element given in \cite{Cirigliano:2011ny}. For the $K_L$ branching ratio,  we take  $\BR(K_L\to \pi^0\pi^0)=8.64\times 10^{-4}$ and $\BR(K_L\to \pi^0\gamma\gamma)=1.29\times 10^{-6}$.  

Starting from $K_L$ momentum and decay vertex, $\{p_{K_L}, Z_{K_L}\}$, the flow of our  simulation for the two body decays is is as follows, 
\begin{align}
\{p_{K_L}, Z_{K_L}\} 
&\xrightarrow[K_L\to \pi^0a/\pi^0]{}&& \{p_{\pi^0}, p_{a/\pi^0}, Z_{K_L}\}
\\
&\xrightarrow[\ \ \pi^0,a\to \gamma\gamma \ \ ]{}&
& \{p_{\gamma_1}, p_{\gamma_2},p_{\gamma_3},  p_{\gamma_4}, Z_{K_L}\}
\\
&\xrightarrow[\vec{p}_\gamma,Z_{K_L}\mapsto (x,y)_\gamma]{}&
&\{(x,y, E)_{\gamma_1}, (x,y, E)_{\gamma_2}, (x,y, E)_{\gamma_3}, (x,y, E)_{\gamma_4} \}.
\end{align}
In the last step, the information of the photon momenta and the $K_L$ decay position is mapped onto the photon positions ($x,y$) on the plane of the ECAL.  
The three body decay $K_L\to \pi^0 \gamma\gamma$ is treated in a similar fashion.

\subsubsection*{Detector's finite resolution}
To take into account the  detector effects, that is the photon's finite energy and position-resolution, we smear each photon's energy and position following the detector resolution~\cite{Sato:2015yqa},\footnote{Before the detector upgrade (see e.g.~\cite{Masuda:2015eta}), the ECAl resolution was 
$
{\sigma_E}/{E}={1.9\%}/{\sqrt{E/\rm GeV}}\oplus 0.6\%  \,.$}
\begin{align}
\frac{\sigma_E}{E}&=\frac{1.74\%}{\sqrt{E/\GeV}} \oplus 0.99\%\,, \label{ECALsmear} \\
{\sigma_{\rm position}}&=2.50{\rm \,mm}\oplus \frac{4.40}{\sqrt{E/\GeV}}{\rm \,mm}\,,
\end{align}
where ${\sigma_{\rm position}}=({\sigma_x\oplus \sigma_y})/{\sqrt{2}}$. 
Taking these as a standard deviation of a gaussian distribution, we smear each photon hit as
\begin{align}
(x,y, E)_{\gamma} \to (x,y, E)^{\rm detect}_{\gamma}\,.
\end{align}
Thus, outputs of our MC samples are $(x,y, E)^{\rm detect}_{\gamma_{1,2,3,4}}$, where the energy smearing dominates the total smearing. 
There are other detector effects, such as photon inefficiency, shower-shape, and timing-resolution, but these are beyond our  simulation setup and the effects are expected to be minor   {because we reproduced shapes and normalizations of several measurements (see Appendix~\ref{sec:validation}). }

\subsection{Event Selection}\label{sec:EVselect}

\subsubsection{Preselection of four photon events}\label{sec:preselect}
Our preselection of four-photon events is similar to the  one used for the four-photon analysis for the $K_L\to\pi^0\pi^0$ decay~\cite{Masuda:2015eta}.  We employ a series of basic cuts on photon energies and positions. 

\begin{enumerate}
\item
The four photons should hit the front ECAL, which is a circle of 1\,m radius. No photons hit the main barrel of the detector (see Fig.~\ref{KOTOExp}). 
\item
As the CsI calorimeter has Moliere radius of about 3.5\,cm for the electromagnetic shower, the four photons are required to be inside a 90\,cm radius, $R_{max}=\max[r_i]<90$cm.   

\item
The position of the innermost photon should be outside of the beam hole, $\max[|x_i|, |y_i|] \geq 7.4\,$cm. 

\item
The photons should be well separated such that $d_{min}=\min |\vec{r}_i-\vec{r}_j| \geq 15\,$cm. This ensures that there are at least four clusters of hits, and, thus, events with four or more photons.

\item 
The minimal energy of each single photon should be $\min [E_{\gamma_i}]\geq 50\,$MeV.

\item 
The total photon energy should be $\sum_i E_{\gamma_i}\geq 350\,$MeV.  

\end{enumerate}
The efficiency of the above preselection cuts is  about 7\% for both signal and background, except the efficiency for $K_L\to \pi^0 a$ with $m_a\leq 20$~MeV that is about 1\% or less due to CUT4 and CUT5 described above. 

\subsubsection{Cuts after reconstruction}\label{Sec:cutsAndEnergy}
After the preselection defined in Sec.~\ref{sec:preselect}, the  search for axion decay to a pair of photons, within the multi photon events proceeds via the following set of cuts\footnote{Wherever relevant we have followed the cut-flow described in~\cite{Masuda:2015eta}.}:
\begin{itemize}
\item[7.]
The reconstructed vertex should be within $2{\rm \,m} < Z_{\rm vtx} < 5.4{\rm \,m}\,,$ which defines the fiducial volume of this analysis. 

\item[8.]
The four-photon invariant mass should match the $K_L$ one, namely, $|m_{4\gamma}-m_{K_L}|<20\,\MeV$.   
\item[9.]
The invariant mass of the photon pair which corresponds to the non-pion candidate, $m_{\gamma_1\gamma_2}$, is required to be away from the neutral pion mass, $|m_{\gamma_1\gamma_2}-m_{\pi^0}|>10~\MeV$. This cut particularly removes most of the $K_L\to \pi^0\pi^0$ background.

\item[10.] 
To further remove the $\pi^0\pi^0$ background, we examine all the possible di-photon pairings to check if any of them reproduces the $K_L\to \pi^0\pi^0$ decay topology. The event is discarded if, for any pair assignment, the pair $A $ satisfies $|m_{\gamma_1\gamma_2}-m_{\pi^0}|<20~\MeV$ and $|m_{4\gamma}-m_{K_L}|<50~\MeV\,$. Only 1.6\% of $\pi^0\pi^0$ background remains after this cut while the other decay topologies are almost unchanged. 
\end{itemize}
When we compute the sensitivity to axion masses near the pion mass, then CUT10 is excluded. This cut would, in fact, substantially reduce the signal for $m_a=(130-140)$~MeV,

The overall efficiencies of CUT1 - CUT10 are $9\times10^{-5}$ and 5\% for $K_L\to \pi^0\pi^0$ and $\pi^0\gamma\gamma$, respectively. The signal efficiency is 3-6\% except for $m_a\lesssim 20~\MeV$ or $m_a\sim m_{\pi^0}$. 
Our background {$\pi^0\pi^0$} MC statistics is poor for $m_{\gamma\gamma}<50$~MeV, we thus treat this region as a single bin for $m_a=1,10,20,30,40~\MeV$.

{So far, our simulation setup does not incorporate the veto cuts and the shower shape cut that are adopted by the KOTO analysis \cite{Ahn:2018mvc}. To take into account the efficiencies of the veto and shower shape cuts, we multiply the kinematic acceptance obtained above by $\epsilon_{\rm veto}=17\%$ and $\epsilon_{\rm shower}=52\%$   \cite{Ahn:2018mvc} for both signal and background\footnote{This is an approximation since the efficiencies in \cite{Ahn:2018mvc} are for the two-photon plus missing energy analysis, while our analysis uses four-photon. We expect that the veto cut efficiency in our analysis can be larger than 17\% because the expected signal does not rely on the missing energy which requires careful veto cuts. At the same time, we expect the efficiency of the shower shape cut to be smaller due to higher multiplicity of photons. 
The precise estimate of the efficiencies requires a full detector simulation which is beyond the scope of this paper. }.}

After all cuts and efficiencies, the remaining background events are mainly from combinatorics of $K_L\to \pi^0\pi^0$ and an irreducible $K_L\to \pi^0\gamma\gamma$. The corresponding distributions are shown Fig.~\ref{Fig:BackSig} in gray and red, respectively. 
In the same plot we also show the expected signal for axion masses of $50\,$MeV and $200\,$MeV respectively and $\BR(K_{\rm L}\rightarrow \pi^0a)=10^{-9}\,.$ {The number of events is computed assuming the future KOTO luminosity of $N_{K_L}^0=2\times 10^{14}$.}

\begin{figure*}[t!] 
\center
\includegraphics[width=11.4cm]{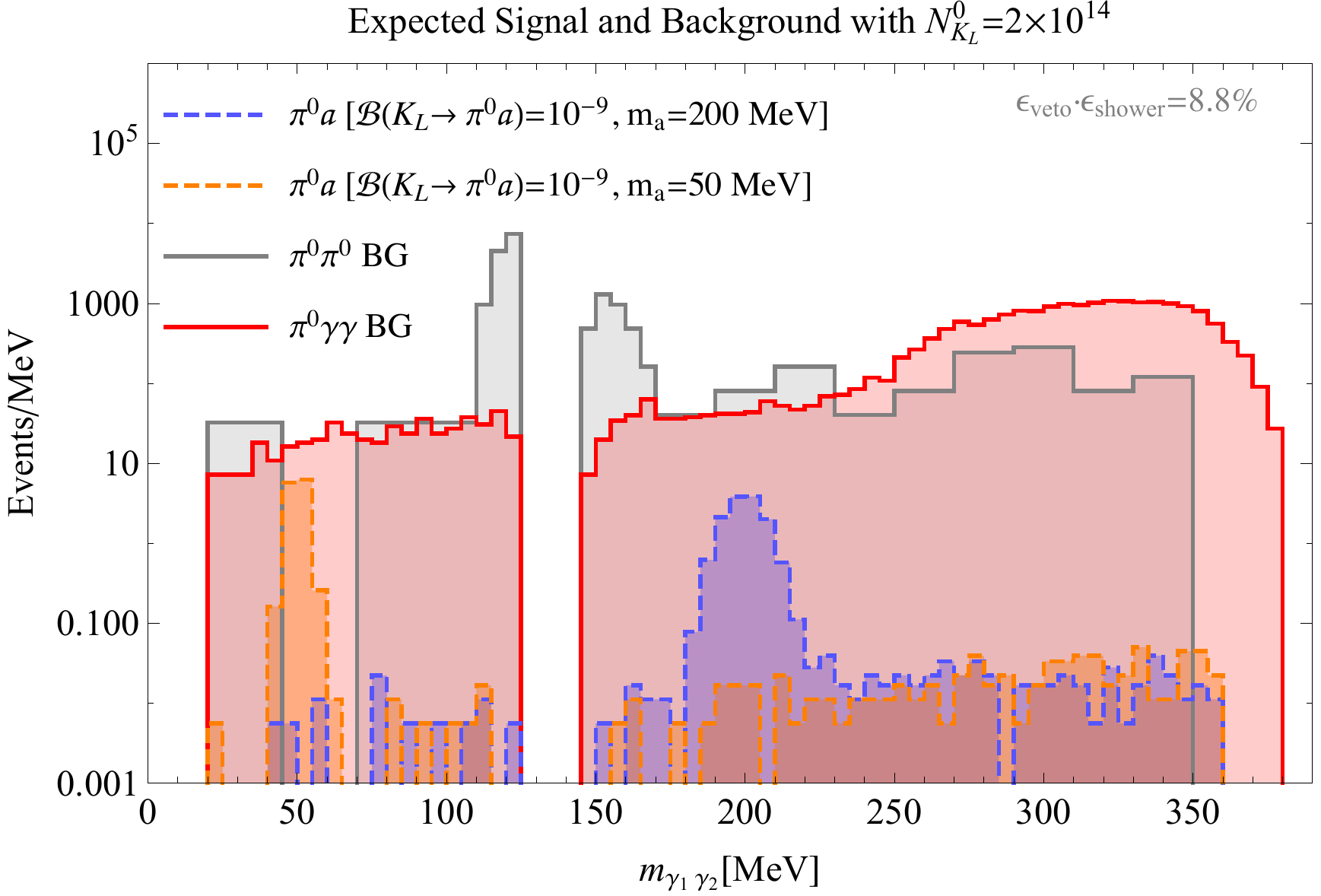}  
\caption{The expected background distribution after cuts with $N^0_{K_L}=2\times 10^{14}$. 
For comparison, the distributions for the signal of a ALP with $m_a=50,~200~\MeV$ and ${\BR}(K_L\to\pi a\to 4\gamma)=10^{-9}$ are also shown. We use a larger bin size for the $\pi^0\pi^0$ background except near the pion mass due to the poor MC statistics of the remaining events. For this figure, we also include the efficiencies of veto and shower cut ($\epsilon_{\rm veto}=17\%$ and $\epsilon_{\rm shower}=52\%$).}
\label{Fig:BackSig}
\end{figure*}

\begin{figure*}[t]\label{fig:resol} 
\center
\includegraphics[width=7.5cm]{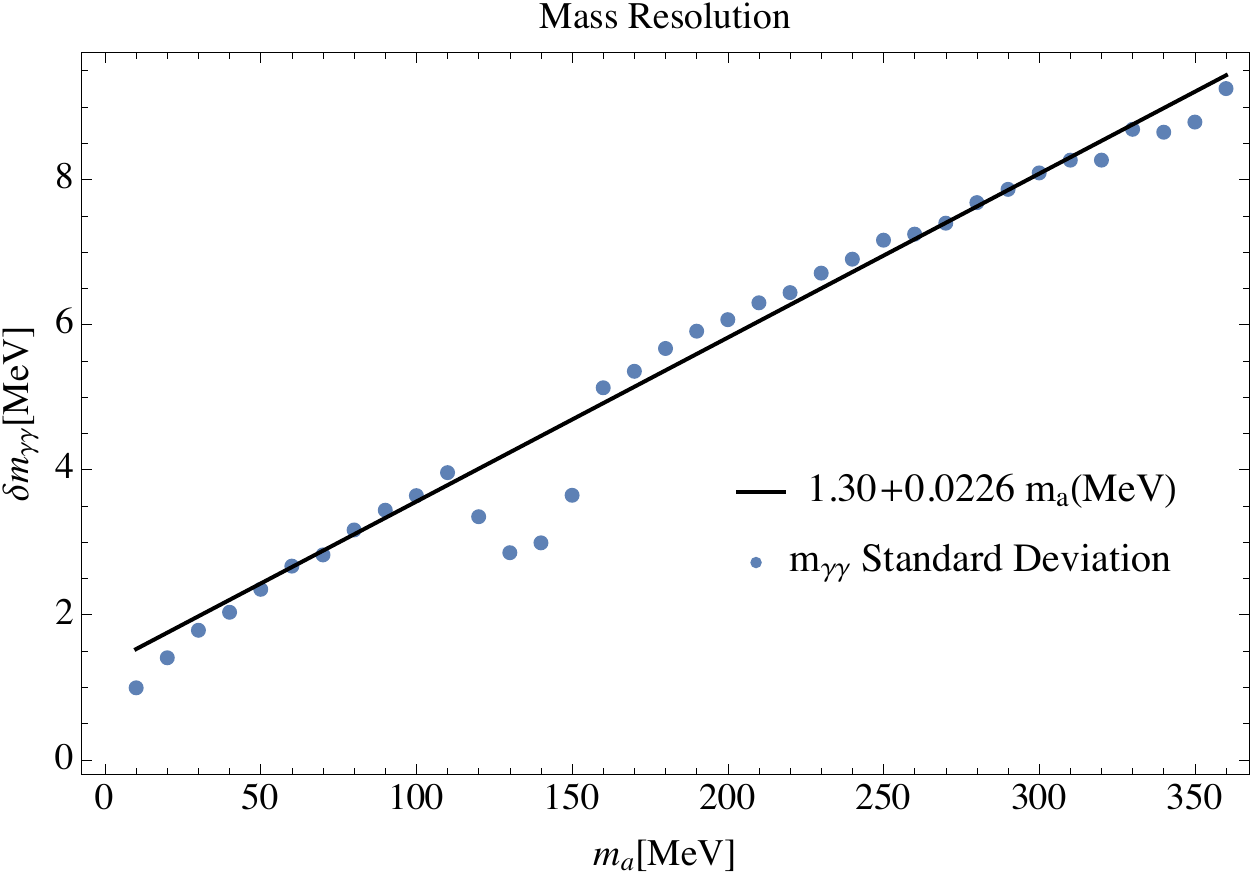}  
\caption{The resolution function and standard deviation in the di-photon invariant mass for $K_L\to \pi^0 a$. In calculating the standard deviation, samples after CUT10 and after the requirement of correct pairing and vertex choice are used. The linear fit is performed excluding the pion mass region, $120~\MeV\leq m_a\leq 150~\MeV$ where CUT10 cannot be applied. }
\label{Fig:vertex}
\end{figure*}

Next we 
estimate the typical size of the ALP di-photon peak. 
Given the above cuts, we have found the RMS of the peak as a function of the mass by fitting the result of our MC simulation to an approximate functional dependence.
By fitting the typical resulting width of the peak around $m_a$ to its RMS value, the peak region is defined as $|m_{\gamma_1\gamma_2}-m_a|<2\delta m_{\gamma\gamma}(m_a)\,,$ with $\delta m_{\gamma\gamma}(m_a)=1.30~\MeV+0.0226m_a$ (see Fig.~\ref{fig:resol}).

Before ending this section, let us briefly discuss another potential source of background: the three-body decay $K_L\to 3\pi^0$. Although this Kaon decay mode has a large branching ratio, $\BR(K_L\to3\pizero)\simeq 0.2$, the photon multiplicity is six. The impact of this background in the $4\gamma$ analysis will crucially depend on the photon inefficiencies of the Main Barrel detector ($\epsilon_\gamma$), which is at the level of $10^{-3}$ \cite{500001058449, talkbyYauWah}. 
We estimate the total efficiency of the $3\pi^0$ background as
\begin{align}\label{Eq3pi}
\BR(K_L\to3\pi^0)\times 30 \times \epsilon_\gamma^2 \times \epsilon_{\rm other} \sim 
6\times 10^{-7} \left(\frac{\epsilon_\gamma}{10^{-3}}\right)^2\left(\frac{\epsilon_{\rm other}}{0.1}\right),  
\end{align}
where the factor of 30 is from combinatorics, and $\epsilon_{\rm other}$ are efficiencies other than photon one. 
Eq. (\ref{Eq3pi}) can be compared to the other physics backgrounds, $K_L\to \pi^0\gamma\gamma$ and $K_L\to 3\pi^0\gamma\gamma$: $\BR(K_L\to\pi^0\gamma\gamma)\times \epsilon_{\pi^0\gamma\gamma}\simeq  8\times 10^{-6}$ and $\BR(K_L\to\pi^0\pi^0)\times \epsilon_{\pi^0\pi^0}\simeq  8\times 10^{-8}$.  
Thus, $K_L\to 3\pi^0$ background is subdominant but can be larger than $K_L\to\pi^0\pi^0$, which may affect the sensitivity at low mass $m_a\lesssim 100~\MeV$ (see Fig.~\ref{Fig:BackSig}). The simulation of $K_L\to3\pi^0$ background requires the full detector simulation, which is beyond the scope of the paper.

\subsection{Displacement and energy of axion decay}
Light ALPs tend to be long-lived because  hadronic final states are kinematically forbidden. Decay with up to  5~cm displacement is effectively prompt decay in our analysis for the KOTO experiment. In fact, the resolution of the reconstructed vertex due to the finite energy resolution of the photons is typically 5~cm. This is shown in  Fig.~\ref{fig:dis}, where a comparison between the location of the true and reconstructed vertex location is shown for different axion masses.

\begin{figure*}[t]\label{fig:dis} 
\center
\includegraphics[width=9.5cm]{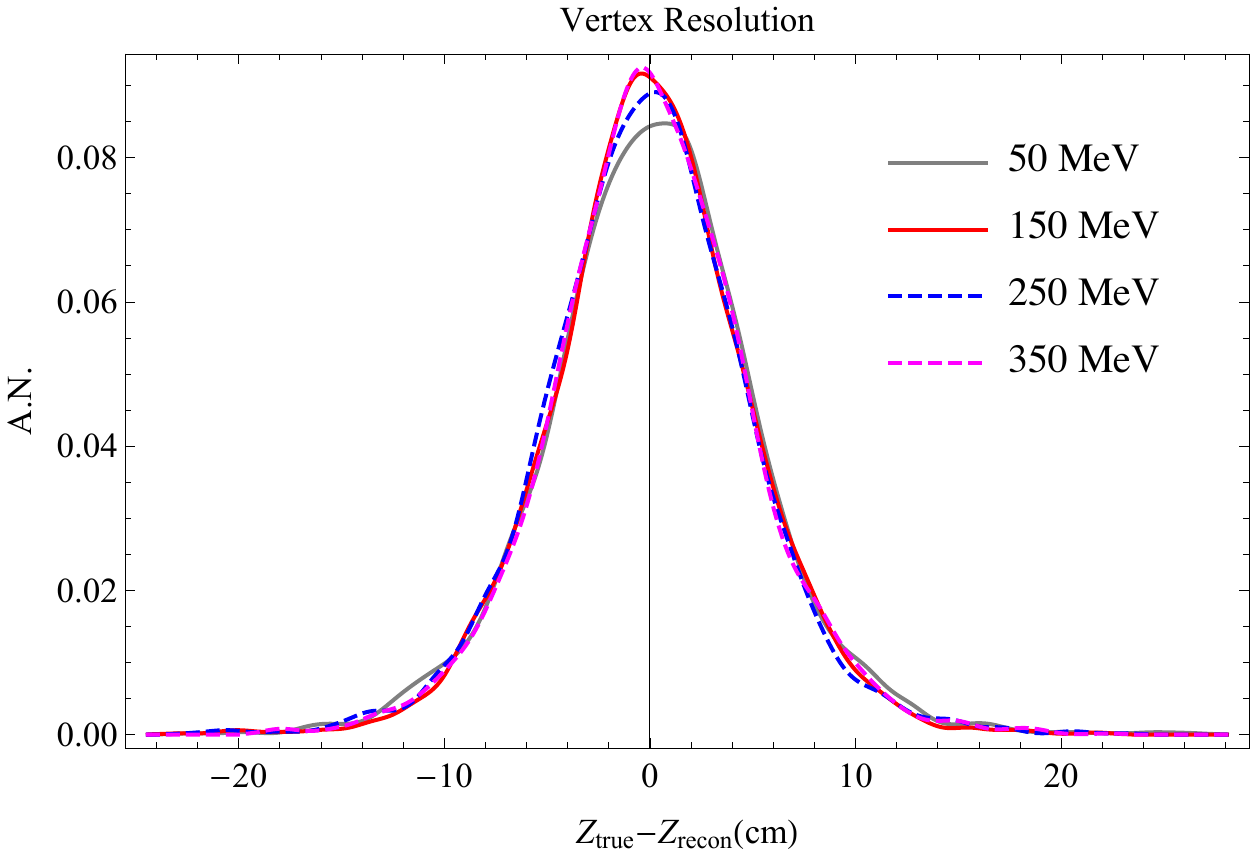}  
\caption{The vertex resolution for the preselected samples of $K_L\to \pi^0 a$ with $m_a=50,150,250,350~\MeV$. The correct pairing is used for the reconstruction. }
\label{Fig:vertex}
\end{figure*}

In order to remove displaced decay with a displacement larger than 5~cm\footnote{Large displacements introduce an extra unknown information rendering the current reconstruction algorithm suboptimal.}, each signal event of our MC simulation is weighted by $(1-\exp[-\frac{5 {\rm cm}}{c\tau_a (E_a/m_a)}])$ where $\tau_a$ is the ALP mean life-time and $(E_a/m_a)$ is the relevant boost factor, $\beta\gamma$. 
{The ALP energy distribution used to compute the boost factor is shown in Fig. \ref{Fig:EnergyALP} for different values of the ALP mass.}

\begin{figure*}[h] \label{fig:Ea}
\center
\includegraphics[width=7.2cm]{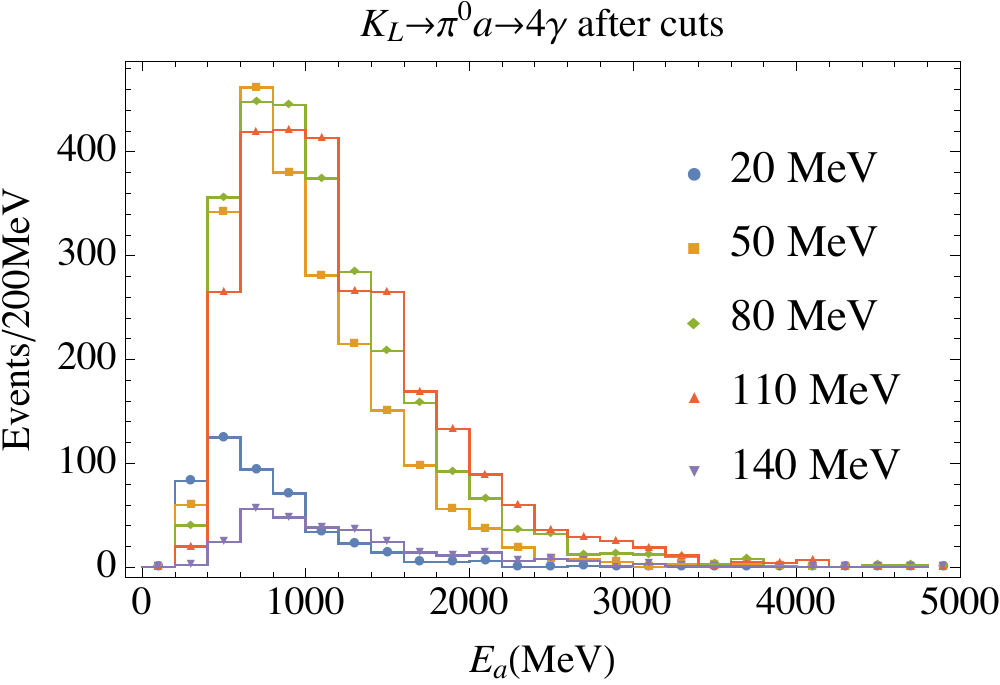} ~~~~ 
\includegraphics[width=7.2cm]{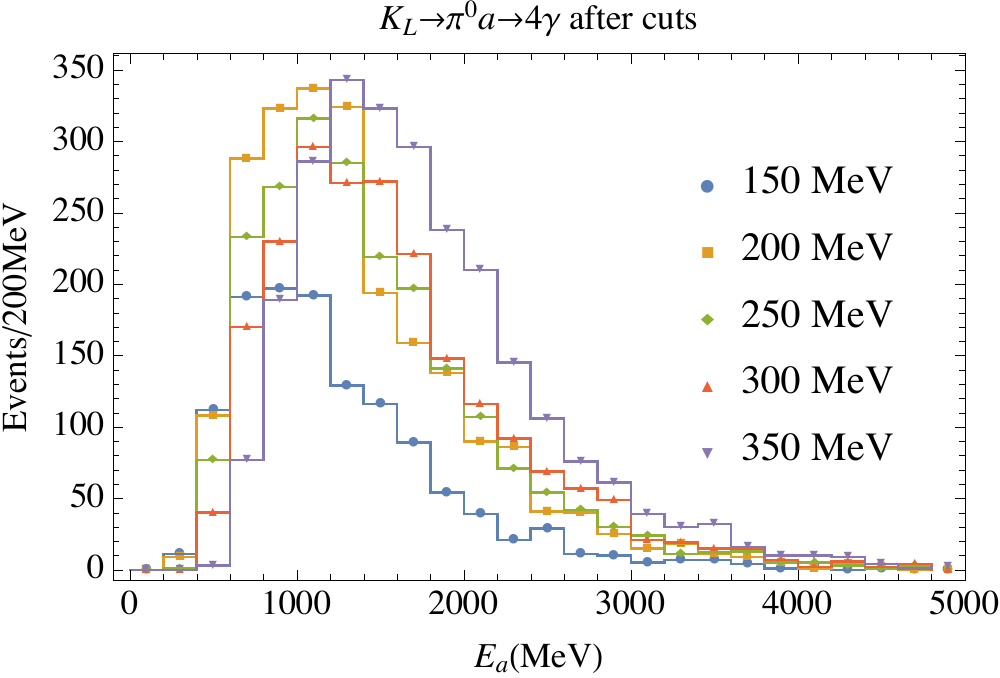}  
\caption{Distribution of the energy of the ALP produced in $K_L\to\pi^0 a$ decays for different ALP masses after cuts. The original MC sample size is $5\times 10^4$, and the different normalization of the several curves indicate the difference in the cut efficiency depending on the ALP mass.}
\label{Fig:EnergyALP}
\end{figure*}

\subsection{Expected sensitivity to the four photon search}\label{sec:Brreach}
The solid blue line in Fig. \ref{Fig:ModelIndependentReach} shows the future reach of the KOTO experiment to the $K_L\to\pi^0 a\to 4\gamma$ signature, as a function of the ALP mass for both $m_a<m_\pi$ and $m_a>m_\pi$. We assume that the sensitivity is determined by the statistical uncertainty, and the decay of the ALP to di-photon is treated as a prompt decay. 
To obtain this curve, we require $S>2\sqrt{B}$ where $S$ is the number of signal events from $K_L\to\pi^0 a$ and $B$ is the number of background events from $K_L\to\pi^0\pi^0, \pi^0\gamma\gamma$. The analysis is done with the future KOTO luminosity of $N_{K_L}^0=2\times 10^{14}$. {From the figure we observe that KOTO can be sensitive to branching ratios as small as few$\times 10^{-9}$.}

{This proposed search can have systematic uncertainties from the determination of the SM background. In  Fig. \ref{Fig:ModelIndependentReach},} we show the cases of 1\% and 10\% systematic uncertainties as  dashed purple and green lines, respectively. 
{We expect these two curves to be very conservative. In fact, the expected signal has a reasonably narrow peak shape, which will allow data-driven background subtraction such as side-band technique. }

\begin{figure*}[h] 
\center
\includegraphics[width=10.cm]{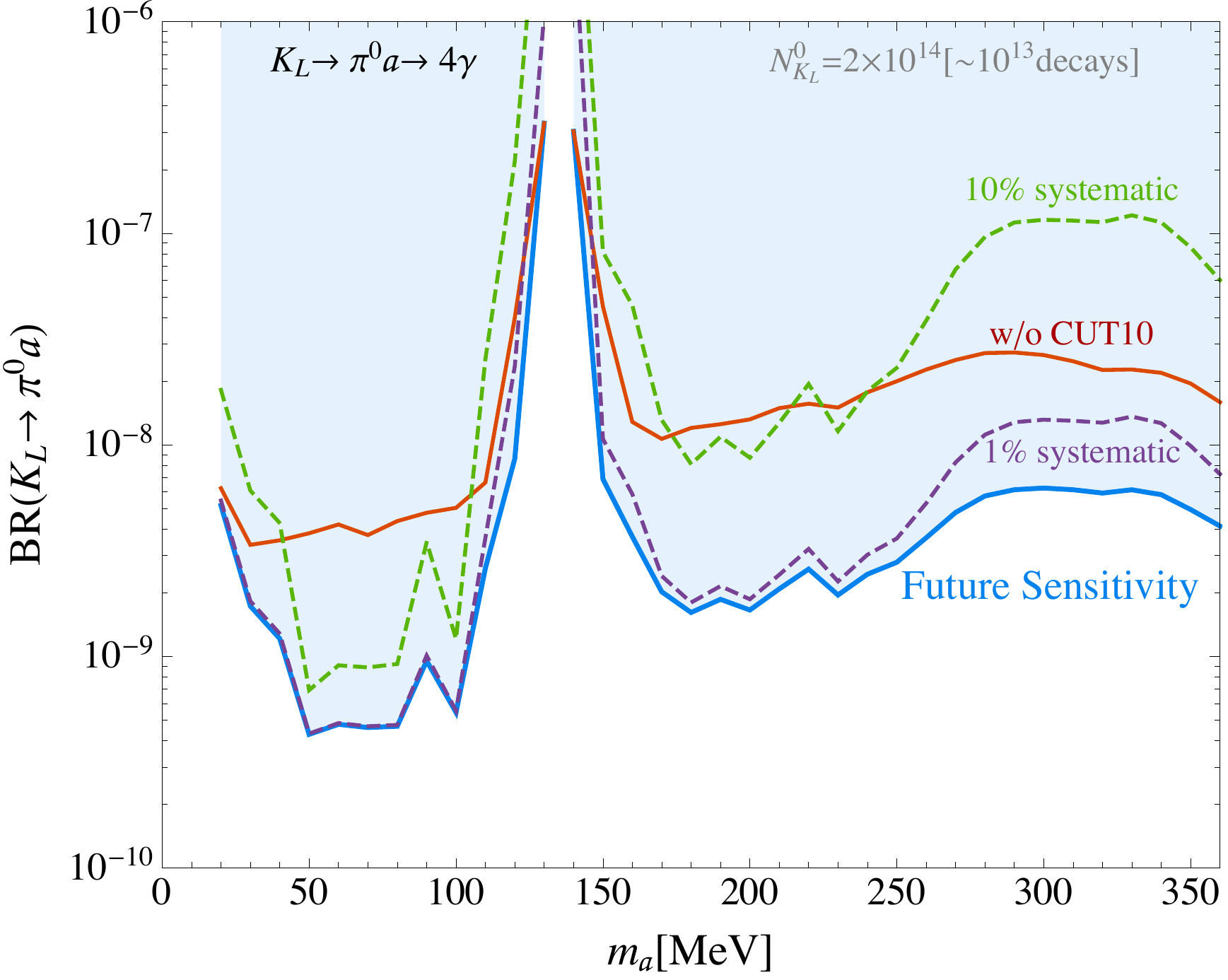}  
\caption{
The future reach of the KOTO experiment to $\BR(K_L\to\pi a\to 4\gamma)$. The blue line shows the 2$\sigma$ sensitivity, {keeping into account only the statistical uncertainty.} The dashed purple line corresponds to the sensitivity with the systematic uncertainty of 1\%, i.e., $S>2(\sqrt{B}\oplus 0.01B)$, and the dashed green  line is for 10\% systematics. Also, the red line corresponds to the sensitivity of the analysis without CUT10 which is effective for ALP masses close to the pion mass, $m_a=130,140~\MeV$. 
}
\label{Fig:ModelIndependentReach}
\end{figure*}

\section{Axion simplified models} \label{sec:simplified}
In this section, we study the KOTO sensitivity to the $K_{\rm L}\rightarrow \pi^0(\gamma\gamma)a(\gamma\gamma)$,  four-photon final state, in terms of several ALP simplified models. We also compare the reach to other past and present high intensity experiments.
\subsection{\boldmath SU(2) coupled axions}\label{Sec:SU(2)}
\subsubsection{Introduction to the model}
We consider a simplified model where the ALP couples only to the field strengths of the $\mathrm{SU}(2)_W$ gauge bosons:
\beq\label{eq:eft}
\mathcal{L}  = (\partial_\mu a)^2 - \frac{1}{2}m_a^2a^2 - \frac{g_{aW}}{4}\,a\,W^a_{\mu\nu}\tilde{W}^{a\mu\nu}\,,
\eeq
 where $W^a_{\mu\nu}$ is the $SU(2)$ field strength tensor, $\tilde W^a_{\alpha\beta}=\frac{1}{2} \epsilon_{\alpha\beta\mu\nu} W^{a,\mu\nu}$, and the $g_{aW}$ coupling is the leading term in the EFT expansion. This coupling is responsible of Kaon decays into ALPs, through W-loop penguin diagrams. In particular, the charged and neutral Kaons will have a decay width \cite{Izaguirre:2016dfi}:
\begin{eqnarray}
\Gamma(K^+\rightarrow \pi^+a) & = &\frac{m_{K^+}^3}{64\pi}\left(1-\frac{m_{\pi^+}^2}{m_{K^+}^2}\right)^2 |g_{asd}|^2\,\lambda^{1/2}_{\pi^+ a}\,, \\ \label{eq:KLtoPIALPWW}
\Gamma(K_{\rm L}\rightarrow \pi^0a) & = &\frac{m_{K_{\rm L}}^3}{64\pi}\left(1-\frac{m_{\pi^0}^2}{m_{K_{\rm L}}^2}\right)^2 \,\mathrm{Im}(g_{asd})^2\,\lambda^{1/2}_{\pi^0 a}\,,
\end{eqnarray}
where  $\lambda_{\pi a} = \left[1-\frac{(m_a+m_\pi)^2}{m_K^2}\right]\left[1-\frac{(m_a-m_\pi)^2}{m_K^2}\right]$. The effective coupling $g_{asd}$ is given by
\beq
g_{asd} \equiv -\frac{3\sqrt{2}G_{\rm F}m_W^2g_{aW}}{16\pi^2}\sum_{\alpha\in c,t}V_{\alpha d}V_{\alpha s}^*f(m_\alpha^2/m_W^2)\,,
\eeq
with the loop function $f(x) \equiv \frac{x\left[1+x(\log x-1)\right]}{(1-x)^2}$. In our numerical analysis, we use the CKM elements as taken from the CKMfitter Group \cite{Charles:2004jd}. In Fig. \ref{Fig:BRKLWAndLifeTime} we show the branching ratio of $K_{\rm L}\rightarrow \pi^0a$ as a function of the ALP mass, as well as of the $g_{aW}$ coupling (gray dashed curves). In this scenario, the branching ratio of $K^+\rightarrow \pi^+a$ is correlated with the $K_L$ one through the isospin relation, $\BR(K^+\rightarrow \pi^+a)/\BR(K_{\rm L}\rightarrow \pi^0a)\sim 1.8\,$.

\begin{figure*}[t!] 
\vspace{0.cm}
\center \includegraphics[width=7.cm]{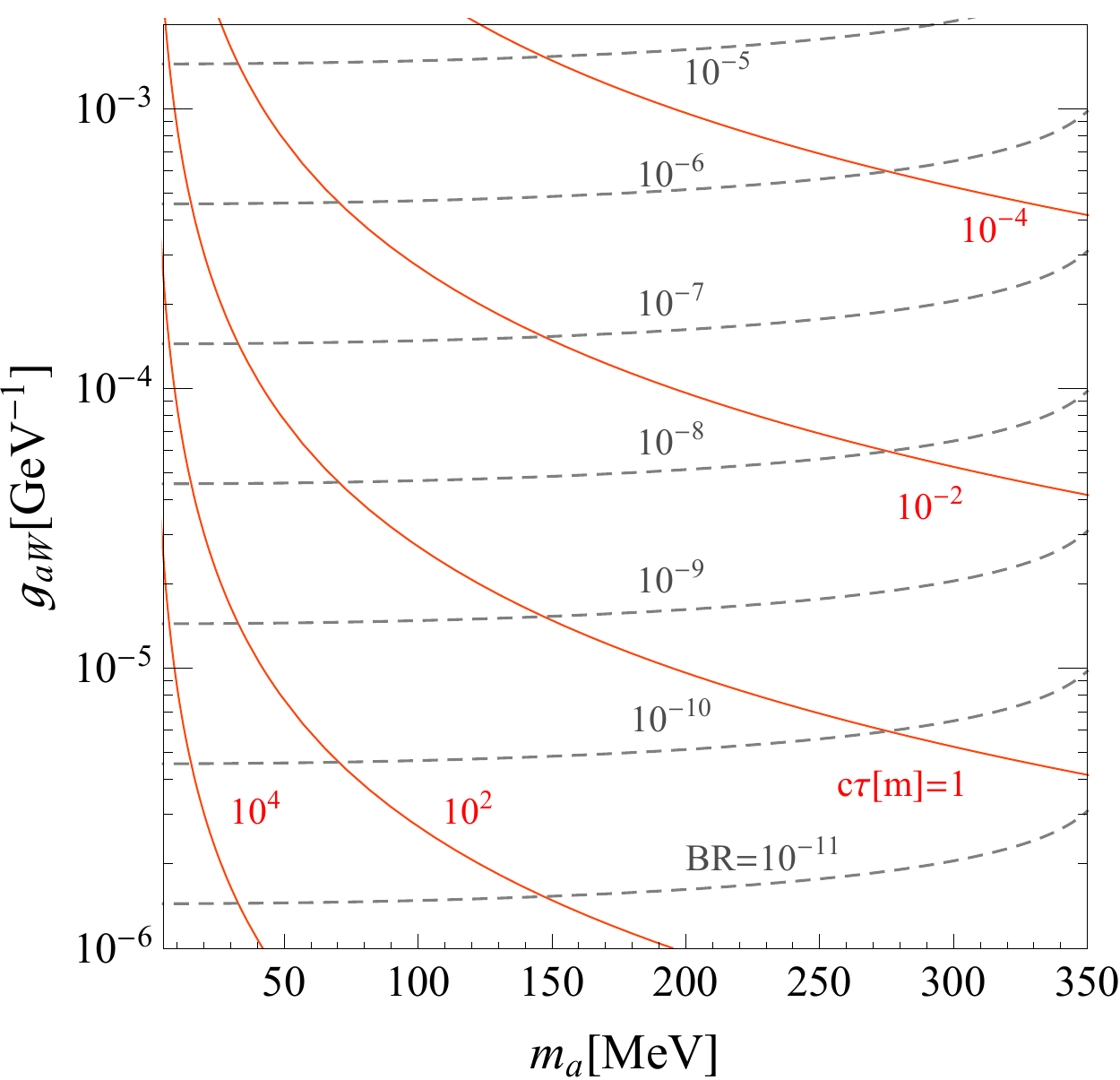}   
\caption{Branching ratio of $K_L\to\pi^0 a$ (in gray, dashed) and proper lifetime of the ALP in meters (in red)  of the SU(2) coupled ALP. The branching ratio of $K^+\to\pi^+ a$ is not shown in the figure since, in this model, it is  simply given by $\sim 1.8\times {\rm{BR}}(K_L\to\pi^0 a)$. 
}\label{Fig:BRKLWAndLifeTime}
\end{figure*}

Once produced, the ALP will decay back to SM particles. In particular, below the pion mass, the axion will decay to photons with a width:
\beq\label{eq:GammaCouplingWW}
\frac{g_{a\gamma\gamma}}{4}aF^{\mu\nu}\tilde F_{\mu\nu}, ~g_{a\gamma\gamma}= g_{aW}\sin^2\theta~~~\Rightarrow~~~ \Gamma(a\to\gamma\gamma)=\frac{g^2_{aW}}{64\pi}\sin^4\theta~ m_a^3\,.
\eeq
In Fig. \ref{Fig:BRKLWAndLifeTime} we show the proper lifetime of the ALP in meters (red curves). 

Similarly, after electroweak symmetry breaking, the ALP will also couple to $Z\gamma$ and $ZZ$. In particular,
\beq\label{eq:CouplingWWModelZ}
\frac{g_{aZ\gamma}}{4}a Z^{\mu\nu}\tilde F_{\mu\nu}, ~g_{aZ\gamma}= g_{aW}\sin\theta\cos\theta;~~~~~~~~\frac{g_{aZZ}}{4}a Z^{\mu\nu}\tilde Z_{\mu\nu}, ~g_{aZZ}= g_{aW}\cos^2\theta\,.
\eeq
As we will discuss in the next section, the former coupling can induce a signal at the LEP experiment, since it induces an exotic decay of the $Z$ boson, $Z\to\gamma a$, with width given by
\beq\label{eq:BRZDecay}
\Gamma(Z\to \gamma a)=\frac{g_{aW}^2\sin^2\theta\cos^2\theta}{96\pi}m_Z^3\,.
\eeq

\subsubsection{KOTO sensitivity and comparison with other experiments}\label{sec:aWWsensitivity}

The KOTO model independent bound presented in Fig. \ref{Fig:ModelIndependentReach} can be interpreted in terms of this ALP simplified model. We compute an ``effective branching ratio'' for $K_L\to\pi^0 a(\gamma\gamma)$ from Eq.~(\ref{eq:KLtoPIALPWW}), taking into account the probability for the ALP to decay within {5} cm from the Kaon decay vertex:

\beq
{\rm{BR}}(K_L\to\pi^0 a,~a\to\gamma\gamma)_{\rm{eff}}={\rm{BR}}(K_L\to\pi^0 a,~a\to\gamma\gamma)\times \left[1-{\rm{exp}}\left(-\frac{5~ {\rm{cm}}}{\tau_a\gamma_a}\right)\right]\,, \label{eq:effectiveBR}
\eeq
where $\tau_a$ is the proper lifetime of the ALP, as shown by the red curves in Fig. \ref{Fig:BRKLWAndLifeTime}. $\gamma_a$ is the boost factor of the ALP that can be easily extracted from Fig. \ref{Fig:EnergyALP}. The corresponding reach is shown in the right panel of Fig. \ref{Fig:BoundWW} by the region delimited by the red dashed line. This bound corresponds to the {``Future Sensitivity''} shown in Fig. \ref{Fig:ModelIndependentReach} for the model independent bound.
The bound is relatively flat above the pion mass and at around $g_{aW}\sim (5-8)\times 10^{-5}/{\rm{GeV}}$. It becomes quite weaker at ALP masses $m_a\lesssim 50$ MeV, because of the weaker bound on the BR$(K_L\to\pi^0 a\to 4\gamma)$ (see Fig. \ref{Fig:ModelIndependentReach}) and because the life time of the ALP becomes quickly macroscopic. 

We now compare this bound to the bounds that we can obtain from other present, past, and future high intensity experiments, {and, in particular, with the NA62 experiment. For additional phenomenological analyses of similar benchmark scenarios, see e.g. \cite{Gavela:2019wzg}.}

\bigskip
\begin{figure*}[t] 
\center
\includegraphics[width=7.2cm]{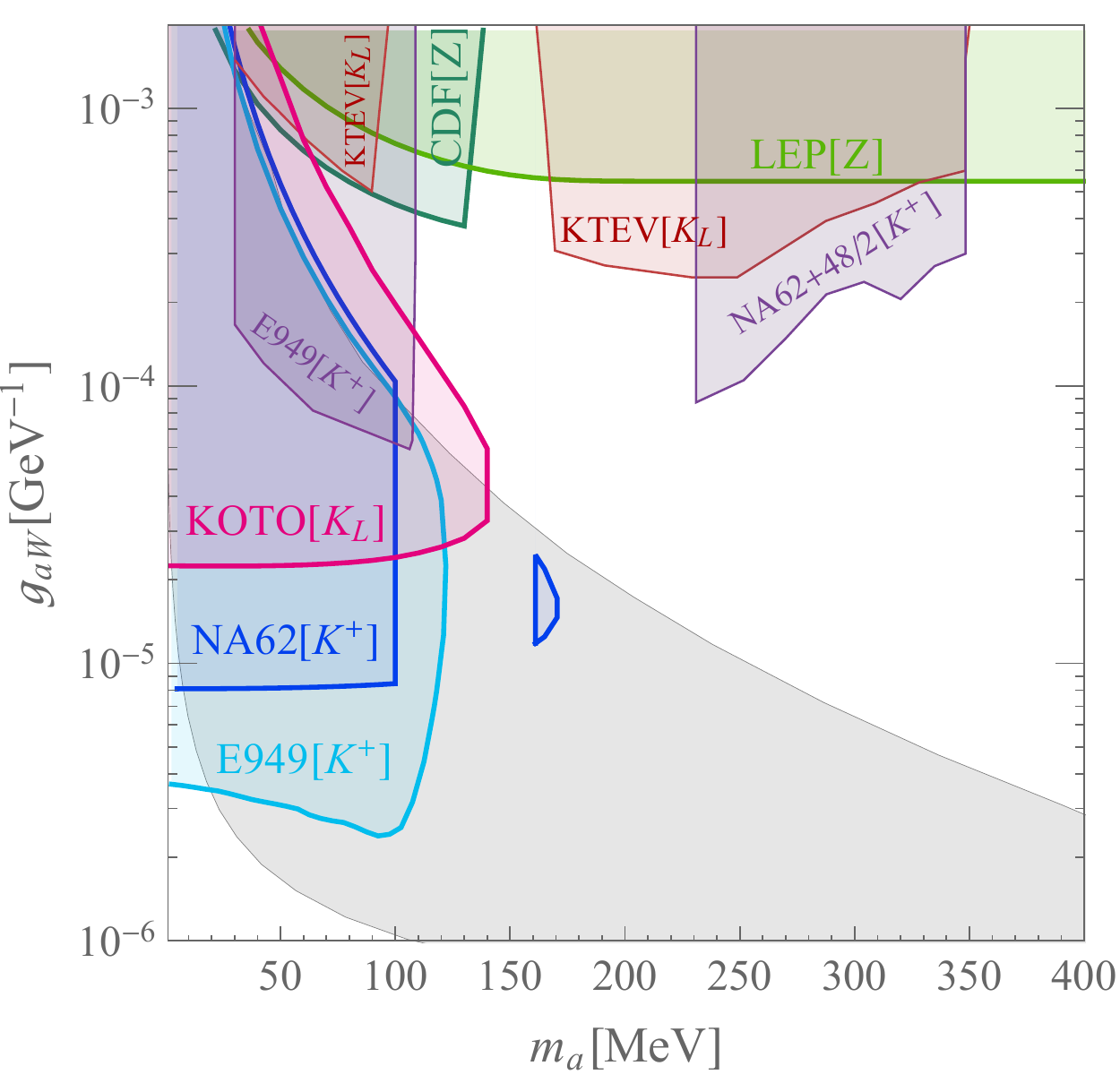} ~~
\includegraphics[width=7.2cm]{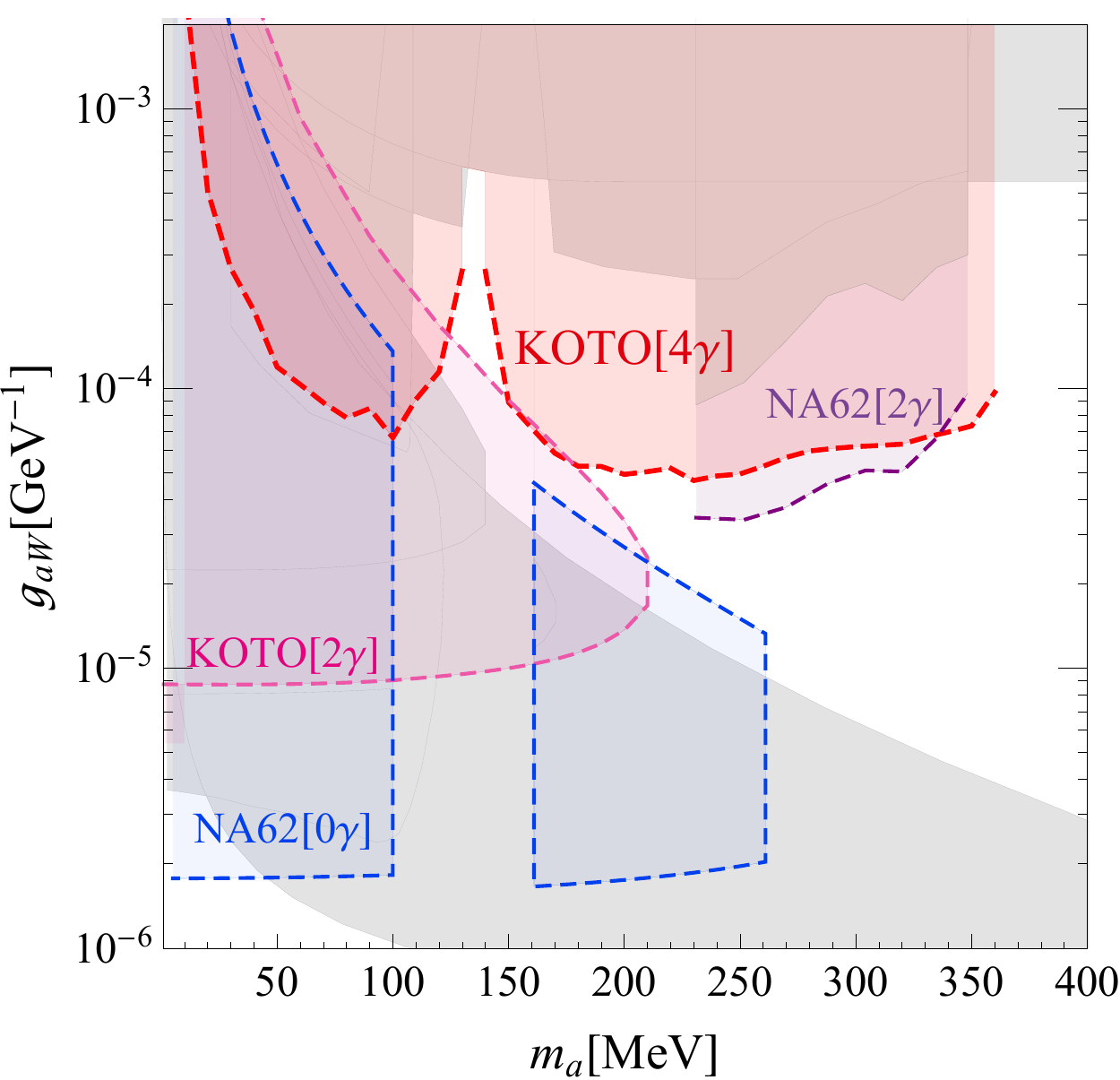} 
\caption{Left panel: Present bounds on the parameter space of the $SU(2)$ coupled-ALPs, as a function of the ALP mass, $m_a$, and of its couplings with $SU(2)$ gauge bosons, $g_{aW}$. Right panel: Present and future bounds on the parameter space. In gray, we present the present bound (as shown in the left panel); in red and magenta, and in purple and blue, we present the future bounds at KOTO ($4\gamma$ and $2\gamma~+$ invisible signatures), and at NA62 ($\pi^++2\gamma$ and $\pi^+~+$ invisible signatures).}
\label{Fig:BoundWW}
\end{figure*}

\noindent
{\bf{Past Kaon experiments}}

Other past experiments looked for an ALP produced from either charged or neutral Kaon decays. The charged Kaon experiments E949 and NA48/2 set an upper bound on the branching ratio of $K^+\to\pi^+\gamma\gamma$ \cite{Artamonov:2005ru,Ceccucci:2014oza} that can be used to set a constraint on a prompt ALP. Similarly, the E949 and NA62 bounds on the SM $K^+\to\pi^+\nu\bar\nu$ decay \cite{Artamonov:2009sz,CortinaGil:2018fkc} can be reinterpreted in terms of a constraint on a long lived ALP. Finally, the KTeV analysis for $K_L\to\pi^0\gamma\gamma$ \cite{Abouzaid:2008xm} can be utilized to set constraints on a prompt ALP, and the KOTO analysis for $K_L\to\pi^0\nu\bar\nu$ \cite{Ahn:2018mvc}  to set constraints on an invisible ALP.

\begin{itemize}
\item {\bf NA48/2, {\boldmath $\pi^+ \gamma\gamma$} analysis}\\
We utilize the NA62/48 measurement of $K^\pm\to \pi^+\gamma\gamma$ in the kinematic range $z=(m_{\gamma\gamma}/m_K)^2>0.2$ \cite{Ceccucci:2014oza} to set a bound on the ALP parameter space. Our analysis is similar to the one done in Ref.~\cite{Izaguirre:2016dfi}, even if we use a different statistical method. In particular, as a conservative bound, we require that the expected signal is less than the observed data plus two sigma uncertainty. We use Fig.~4 of \cite{Ceccucci:2014oza} to set the bound on the branching ratio as a function of the ALP mass for $m_a\in(220-350)$ MeV. We require that the ALP decays in the detector volume, and, more specifically, that the decay length in the lab frame is less than 10 m. We include the corresponding weight factor $(1-\exp[-\frac{10~{\rm m}}{\tau_a (E_a/m_a)}])$ where $E_a$ is taken to be $37~\GeV$ (i.e. half of the Kaon energy). Our bound is shown in violet in the left panel of Fig. \ref{Fig:BoundWW}.

\item {\bf E949, {\boldmath$\pi^+ \gamma\gamma$} analysis}\\
The E949 experiment searched for $K^+$ decays at rest with a pion momentum $p_{\pi^+}>213$ MeV. This analysis was re-interpreted in terms of $K^+\to\pi^+ a,~a\to\gamma\gamma$ with the ALP decaying within 80 cm of the stopped Kaon \cite{Izaguirre:2016dfi}. The corresponding bound is shown in purple at $m_a< 110$ MeV in the left panel of Fig. \ref{Fig:BoundWW}. 


\item {\bf NA62, {\boldmath $\pi^++$ invisible} analysis}\\
In the $K^+\to \pi^+ \nu\bar\nu$ analysis \cite{CortinaGil:2018fkc}, there are two distinct signal regions at low and at high missing mass: $0~\MeV< m_{\rm miss}<100~\MeV$ (R1) and $161~\MeV<m_{\rm miss}<261~\MeV$ (R2), respectively. We calculate the acceptance of $K^+ \to \pi^+a(\rm invisible)$, $A_{\pi^+a}=$ 5.2\% and 7.0\% in R1 and R2 signal regions, respectively~\footnote{We reweigh the reported acceptance for $K^+\to \pi^+\nu\bar\nu$ by the phase-space factor, 
\begin{align}
A_{\pi^+a}^{\rm R1(R2)}[0(161)<m_a/\MeV<100(261)]]=\frac{A_{\pi^+\nu\bar\nu}^{\rm R1(R2)}\cdot \Gamma_{\pi^+\nu\bar\nu}} {\Gamma_{\pi^+\nu\bar\nu}[0(161)<m_{\nu\bar\nu}/\MeV<100(261)]}
=5.2\%(7.0\%)
\end{align}
}. In addition, to compute the total yield, we adopt the same trigger efficiency, $\epsilon_{\rm trigg}=0.90$, and veto efficiency, $\epsilon_{\rm veto}=0.76$ as for the SM decay. The decay $K^+\to\pi^+ a$ is effectively a $K^+\to\pi^++$ invisible decay at NA62, as long as the ALP has a decay length of at least 150 m. Therefore, we compute the number of signal events in the two signal regions as $N_{K^+}\times\rm{BR}(K^+\to\pi^+ a)\times A_{\pi^+a}\times \epsilon_{\rm veto}\times\epsilon_{\rm trigg}\times e^{-150~{\rm{m}}/\tau\langle\gamma\rangle}$, where $N_{K^+}$ is the number of Kaons decaying in the fiducial region ($N_{K^+}\sim1.2\times 10^{11}$ with the present dataset). The mean boost, $\langle\gamma\rangle$, is calculated for each ALP mass using our signal Montecarlo events that pass the cuts on the geometrical acceptance for the charged pion, assuming that all Kaons are produced with an energy of exactly 75 GeV. Based on the NA62 observed number of events, we require the number of signal events to be less than 3.0 in the R1 signal region and less than 4.74 in the R2 signal region. The corresponding bound is shown in blue in the left panel of Fig. \ref{Fig:BoundWW}.


\item  {\bf E949, {\boldmath $\pi^++$ invisible} analysis}\\
The E949 collaboration has interpreted their analysis for the SM $K^+\to\pi^+\nu\bar\nu$ decay in terms of a bound on a new stable massive particle produced from $K^+\to\pi^+X$ \cite{Artamonov:2009sz}. We utilize this result to set a bound on our ALP parameter space. We require that the effective branching ratio for $K^+\to\pi^+ a$ is smaller than the one presented in Fig. 18 of \cite{Artamonov:2009sz}. To compute this effective branching ratio, we compute the probability for the ALP to escape the detector, i.e. to have a life-time longer than 1.5m, starting from a Kaon decaying at rest \footnote{We have verified that the requirement of a life-time longer than 1.5m reproduces the results in Fig.18 of the E949 paper \cite{Artamonov:2009sz} in the case of a finite life-time of X.}. Our bound is presented in cyan in the left panel of Fig. \ref{Fig:BoundWW}.

 \item {\bf KTeV, {\boldmath $\pi^0 \gamma\gamma$} analysis}\\
The KTeV analysis for $K_L\to\pi^0\gamma\gamma$ \cite{Abouzaid:2008xm} has been utilized to set a bound on a prompt (i.e. decaying within 1 m from the $K_L$ decay) ALP decaying into two photons \cite{Izaguirre:2016dfi}. The corresponding bound is shown in red in the left panel of Fig.~\ref{Fig:BoundWW}. 

\item {\bf KOTO, {\boldmath $\pi^0+$ invisible} analysis}\\
The KOTO analysis \cite{Ahn:2018mvc} sets an upper bound on the BR$(K_L\to\pi^0 X)$ where $X$ is an invisible NP particle with mass below $\sim$260 MeV. The analysis utilizes $3.68\times 10^{11}$ $K_L$ decaying inside the detector (see Table~\ref{flux}). Branching ratios as small as $\sim 2.2\times 10^{-9}$ have been tested under the assumption of a $100\%$ invisible decay. We reinterpret this search in terms of our ALP model, requiring that the ALP has a lifetime long enough to decay after the detector. We obtain the distribution of the energy and decay point of $K_L$ based on our MC simulation with the analysis defined in \cite{Ahn:2018mvc}. Our bound is shown in pink in the left panel of Fig. \ref{Fig:BoundWW}.

\end{itemize}

\noindent
{\bf{Past Colliders and beam dumps}}

In addition to Kaon experiments, the LEP and Tevatron colliders also set constraints on the parameter space of this model. This is shown by the two green regions in the left panel of Fig. \ref{Fig:BoundWW}. 

Because of the ALP $g_{aZ\gamma}$ coupling (see Eq. (\ref{eq:CouplingWWModelZ})), the $Z$ boson can decay to $a\gamma$, inducing a multi-photon signature in the LEP detectors \cite{Mimasu:2014nea,Jaeckel:2015jla}. The total and differential cross sections for the process $e^+e^-\to\gamma\gamma$ was measured by the L3 collaboration at around $\sqrt s=91$ GeV \cite{Acciarri:1995gy}. In particular, the L3 experiment set the bound BR$(Z\to\gamma\gamma)<5.2\times 10^{-5}$. This bound is directly applicable to our model at light ALP masses, since the photons from the ALP decay would be collimated in the L3 detector. The dark green region in left panel of Fig.~\ref{Fig:BoundWW} is the bound we obtain from this branching ratio, asking the ALP to decay before the L3 ECAL (and therefore with a decay length smaller than $\sim 0.5$m \cite{Gataullin:2006fv}).  

Similarly, the CDF collaboration searched for the decay of a $Z$ boson into two photons \cite{Aaltonen:2013mfa}. In particular, the collaboration set a bound BR$(Z\to\gamma\gamma)<1.46\times 10^{-5}$ and BR$(Z\to\gamma\pi^0)<2.01\times 10^{-5}$, with the pion detected as a single photon. We apply this more conservative bound on the decay into a photon and pion for ALP masses below the pion mass. To obtain the corresponding bound, we require the ALP to decay before the CDF central electromagnetic calorimeter located at 6.8\,in from the collision point~\cite{Balka:1987ty}\footnote{This type of analysis was done in  \cite{Bauer:2017ris}, with a more conservative bound up to  $m_a<73$ MeV to guarantee collimation of the two photons from the ALP decay.}.  The bound is represented by the dark green region at $m_a<m_\pi$, in left panel of Fig. \ref{Fig:BoundWW}. 

Finally, past electron and proton beam dump experiments set a bound on the coupling of the ALP with photons $g_{a\gamma\gamma}$ (see Eq.~(\ref{eq:GammaCouplingWW})). We take these bounds from \cite{Dolan:2017osp}. They are represented in gray in the left panel of Fig.~\ref{Fig:BoundWW}.

\bigskip
\noindent
{\bf{Future measurements at Kaon experiments}}

Next, we compare  the future sensitivity of KOTO to the four-photon final state (see red region in the right panel of Fig. \ref{Fig:BoundWW}) to other projection of searches of NA62 and KOTO.

In particular, the purple region in the figure represents our projection of the NA48/62 $K^+\to\pi^+\gamma\gamma$ analysis. To produce this region, we scale the NA48/62 $K^+\to\pi^+\gamma\gamma$ {uncertainty} by the $\sqrt L$ where $L$ is the ratio of NA62 and NA48/62 number of Kaon decaying in the fiducial volume. The NA62/48 have used $1.59\times 10^{9}$ $K^\pm$ decays in the fiducial volume \cite{Ceccucci:2014oza}, while for the future projection we assume that the NA62 will collect $10^{13}$ $K^+$ decays  with a downscaling trigger factor of 400 for $K^+\to \pi^+\gamma\gamma$  \cite{CortinaGil:2018fkc}.

The blue region in the figure represents the projection of the NA62 $\pi^+ +$invisible bound utilizing the full future luminosity. The bound corresponds to 12 events obtained with $10^{13}$ $K^+$ decaying in the fiducial region\footnote{We have obtained 12 events via rescaling the number of SM single event sensitivity (0.267) and background (0.152) events observed now by NA62 with $1.2\times 10^{11}$ $K^+$~\cite{CortinaGil:2018fkc}.}.

Finally, the pink region in the figure represents the projected bound for the KOTO $K_L\to\pi^0+$ invisible analysis. To obtain this bound, we scale the bound on the branching ratio in \cite{Ahn:2018mvc} with the $\sqrt L$ ($ L=1.6\times 10^{13}/3.68\times 10^{11}$, see Table \ref{flux}). %

\subsection{\boldmath Gluon coupled axions}~\label{subsec:GG}\vspace{-10mm}

\subsubsection{Introduction to the model}
The axion solution to the strong CP problem makes benchmark scenarios with an ALP coupled to gluons particularly interesting. The effective Lagrangian at the low energy scale $\mu\sim m_a$\footnote{In Appendix \ref{Appendix E}, we will briefly discuss additional UV contributions that can affect the $K\to\pi a$ rate if this effective Lagrangian is, instead, valid at a higher energy scale.}, can be written as 
\begin{align}
\label{eq:eftGG}
\mathcal{L} & \supset (\partial_\mu a)^2 - \frac{1}{2}m_a^2a^2 - \frac{g_{ag}}{4}\,a\,G^a_{\mu\nu}\tilde{G}^{a\mu\nu}
  = (\partial_\mu a)^2 - \frac{1}{2}m_a^2a^2 - \frac{\alpha_s}{8\pi F_a}\,a\,G^a_{\mu\nu}\tilde{G}^{a\mu\nu},
\end{align}%
where $F_a$ is the ALP decay constant and $\tilde G_{\mu\nu}^a=\frac{1}{2}\epsilon^{\alpha\beta\mu\nu}G_{\alpha\beta}^a$.  {Since the effective theory can be valid up to a scale  $4\pi/g_{ag}=4\pi(2\pi F_a/\alpha_s)\sim 4{\rm TeV}({F_a}/{10\,\GeV})$ (while the cutoff can be $4\pi F_a$ in many models), we focus on the ALP phenomenology and ignore the bounds from heavy states.} 

To obtain the form of the effective theory below the $\Lambda_{\rm QCD}$ scale we resort to chiral perturbation theory. For convenience, we perform a chiral rotation of light quarks to remove the $aG\tilde{G}$ coupling \cite{Georgi:1986df} (see also \cite{Kim:2008hd,Bauer:2017ris}) and generate the derivative couplings with the three light quarks (up, down, and strange) at leading order in the chiral Lagrangian: $- \kappa_q(\partial_\mu a/2 F_a) \bar q \gamma^\mu \gamma_5 q $ where $\kappa_q\equiv m_q^{-1}/(m_u^{-1}+m_d^{-1}+m_s^{-1})$. {In our analysis, we also keep the strange quark, since,} as we further discuss below, the mixing with the $\eta$ meson also plays an  important role when computing  Kaon to ALP decay processes~\cite{Bardeen:1986yb}. 

These derivative couplings induce a kinetic mixing between the SM mesons and the ALP. The $\pi^0$ and $\eta$ states receive a small admixture of the physical ALP state, such that %
\begin{align}
\pi^0\simeq \pi^0_{\rm{phys}}+\theta_{\pi a}a_{\rm{phys}}\,, \quad 
\eta\simeq \eta_{\rm{phys}}+\theta_{\eta a}a_{\rm{phys}}\,,
\end{align}
where, at the leading order, the mixing angles are given by
\begin{align}
&{\theta_{\pi a}\simeq \frac{F_{\pi}}{2F_a}
 (\kappa_u-\kappa_d)\frac{m_a^2}{m_a^2-m_{\pi^0}^2}}\,,
\quad 
\\&\label{eq:ThetaALPEta}
{\theta_{\eta a}\simeq \frac{F_{\pi}}{F_a}
\frac{\sqrt{2}m_a^2[\kappa_u+\kappa_d-2\kappa_s]\cos\theta_{\eta\eta'}-2\left(m_a^2[\kappa_u+\kappa_d+\kappa_s]-6\Delta m_\pizero^2\right)\sin\theta_{\eta\eta'})}{2\sqrt{6}(m_a^2-m_\eta^2)}}
\,,
\end{align}
where we have defined $\Delta^{-1}\equiv{(m_u+m_d)(m_u^{-1}+m_d^{-1} +m_s^{-1})}$, and $F_\pi$ is the pion decay constant given by $F_\pi\approx 93$ MeV. $\theta_{\eta\eta'}$ is the $\eta$-$\eta'$ mixing, whose value has a large uncertainty and lies in the range $\simeq -(10^{\circ}$-$20^{\circ})$ (see e.g.~\cite{Christ:2010dd,Guo:2015xva,Pham:2016nla}). Note the different $m_a$ dependence in the ALP-$\eta$ mixing of the $\cos\theta_{\eta\eta^\prime}$ and $\sin\theta_{\eta\eta^\prime}$ terms. This is due to the fact that the $\sin\theta_{\eta\eta^\prime}$ term arises from mass mixing, the $\cos\theta_{\eta\eta^\prime}$ from kinetic mixing.
At the same order in the chiral Lagrangian, the physical masses of the ALP, pion, and eta mesons are unaffected.

\begin{figure*}[t!] 
\vspace{0.cm}
\center \includegraphics[width=8.5cm]{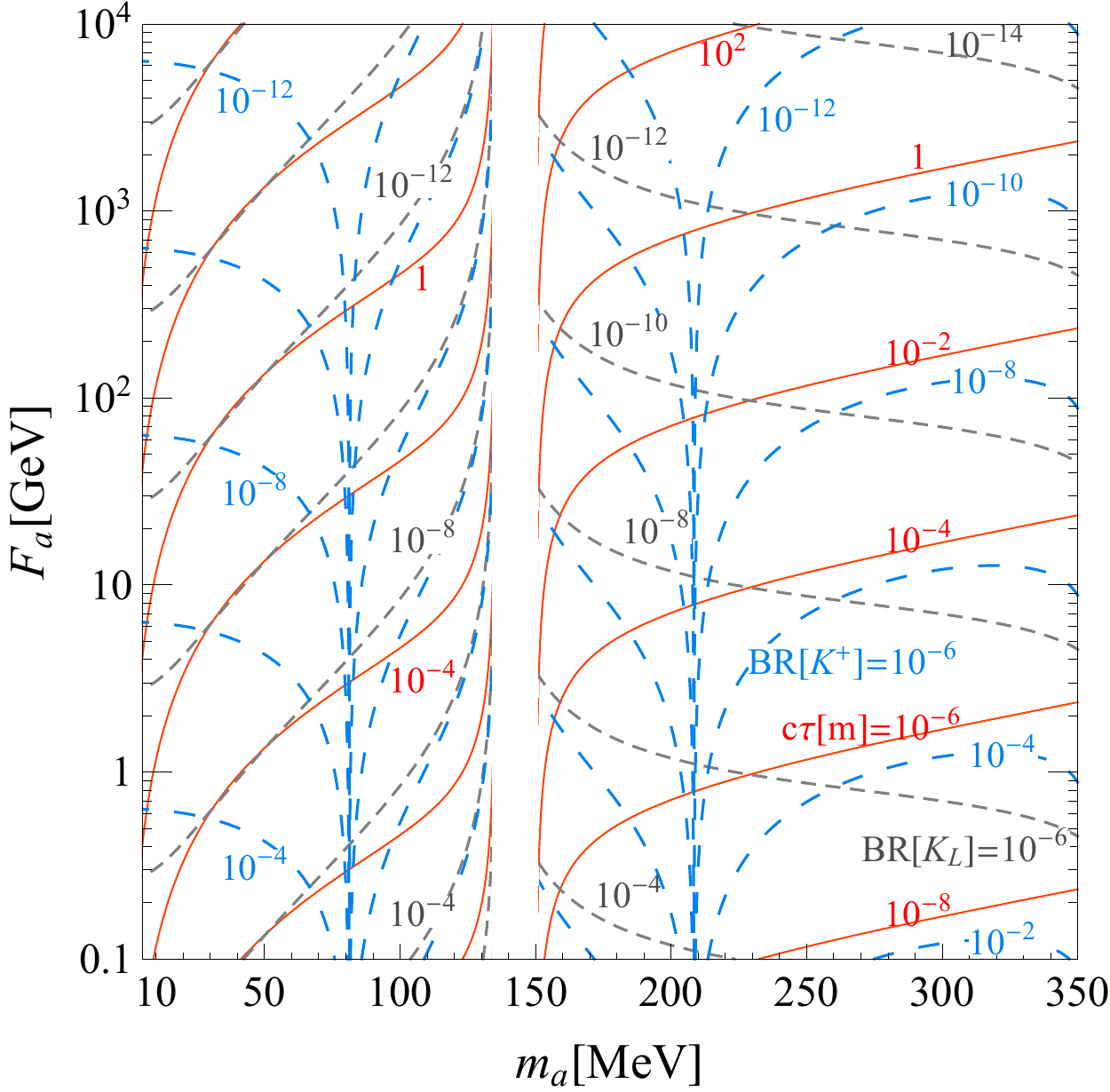}   
\caption{Branching ratio of $K_L\to\pi^0 a$ (in black dashed), branching ratio of $K^+\to\pi^+ a$ (in light blue, dashed) and proper lifetime of the ALP in meters (in red) of the $G\tilde{G}$ coupled ALP. 
The mass range $\sim (135-150)$ MeV is not plotted for a better illustration. 
}\label{Fig:BRKLGAndLifeTime}
\end{figure*}

From the ALP mixing with neutral light mesons and the known operators for hadronic decays of the Kaons in the chiral Lagrangian (see Appendix \ref{Appendix:ALPglue}), we can calculate the Kaon decay widths at the leading order (similar calculations can be found in \cite{Alves:2017avw}). For simplicity, in the following we will fix $\sin\theta_{\eta\eta^\prime}=-1/3$ \cite{Aloni:2018vki}. We will comment in the text, how the results will change if we had fixed a different value of $\theta_{\eta\eta^\prime}$ in the $-(10^{\circ}$-$20^{\circ})$ range.
\begin{align}
&\Gamma(K^+\to \pip a)=\frac{1}{8\pi} |g_{K^+ \pi^- a}|^2\frac{|\vec{p}_a|}{m_K^2}\,, \\\label{eq:KLDecayGG}
&\Gamma(K_L \to \pizero a)=\frac{1}{8\pi} |\sqrt{2} \epsilon_K g_{K^0 \pi^0 a}|^2\frac{|\vec{p}_a|}{m_K^2}\,,
\end{align}
 where the CP violating parameter in the Kaon mixing is given by $\epsilon_K=2.23\times 10^{-3}$, {and $|\vec{p}_a|$ is the absolute value of the momentum of the ALP.} The corresponding effective couplings are
\begin{align}\label{gKpia}
g_{K^+ \pi^- a} \nonumber
= -i F_{\pi} \Bigg\{&
 \theta_{\pi a}\frac{ 3 G_8 \left(m_a^2-m_\pip^2\right)+G_{27} \left(5 m_{K^+}^2-7 m_\pip^2+2
   m_a^2\right)}{3 }\,
   \nonumber\\
   +&
   \theta_{\eta a} \frac{6G_8
   \left( m_{K^+}^2- m_a^2\right)
   +G_{27}
   \left(7 m_{K^+}^2-3m_\pip^2-4 m_a^2\right)}{3\sqrt{3}}
   \Bigg\}\,,\\
   %
   g_{K^0 \pi^0 a} \nonumber
   =-i F_{\pi}
   \Bigg\{&
   \theta_{\pi a}\frac{  (G_8-G_{27}) \left(2
   m_{K^0}^2-m_\pizero^2-m_a^2\right)}{\sqrt{2}}
   \nonumber\\
 +& \theta_{\eta a}\frac{ 
   2G_8 \left(- m_{K^0}^2+m_a^2\right)+G_{27}
   \left( m_{K^0}^2+m_\pizero^2-2 m_a^2\right)}{ \sqrt{6}}\,
   \Bigg\}.
\end{align}

The $G_8$ and $G_{27}$ couplings are the coefficients in front of the operators responsible for the $\bar s\to\bar d$ transition, which transform like $(8_L,1_R)$ and $(27_L,1_R)$ (see Appendix \ref{Appendix:ALPglue}). From lattice calculations, we know that the $G_8$ coefficient is significantly larger than $G_{27}$. In our numerical analysis we will use the leading order values~\cite{Cirigliano:2011ny},  
\begin{align}
&G_{8,27}=-\frac{G_F}{\sqrt{2}}V_{ud}V^*_{us}~g_{8,27}\simeq -\frac{1.80\times 10^{-6}}{\rm GeV^2} g_{8,27},  \nonumber\\
&g_8=4.99, \quad g_{27}=0.253, \label{eq:G8G27}
\end{align}
where $V_{ud},V_{us}$ are CKM elements.

In Fig. \ref{Fig:BRKLGAndLifeTime}, we show the BR$(K_L\to\pi^0a)$ and BR$(K^+\to\pi^+a)$ as a function of $m_a$ and of the decay constant $F_a$. As we expect from the $\epsilon_K$ suppression in Eq. (\ref{eq:KLDecayGG}), the branching ratio of the neutral mode is generically suppressed, if compared to the one of the charged mode. There are also some accidental cancellations of the charged Kaon branching ratio. The position of the cancellation at low mass $m_a\sim 80$ MeV largely depend on the particular value chosen for $\theta_{\eta\eta^\prime}$, whose uncertainty is sizable. The position of the cancellation at higher mass $m_a\sim 210$ MeV, instead, depend importantly on both the exact values of the quark masses, and the mixing angle $\theta_{\eta\eta^\prime}$.

The decay of the ALP is controlled by the di-photon coupling and it 
is generated by the chiral rotation and the mixing with the mesons,
\begin{align}
&\frac{g_{a\gamma\gamma}}{4}a F_{\mu\nu}{\tilde F^{\mu\nu}}
,\nonumber
\\
&\quad g_{a\gamma\gamma}=
 \frac{\alpha N_c}{\pi}\left(-\frac{1}{F_a} {\rm tr}[\tilde\kappa_q Q_q^2] 
 +\frac{\sqrt 2}{F_{\pi}} {\rm tr}[\lambda_3 Q_q^2]  \theta_{a\pi}+\frac{\sqrt 2}{F_{\pi}} {\rm tr}[(\lambda_8\cos\theta_{\eta\eta'}-\lambda_0 \sin\theta_{\eta\eta'}) Q_q^2]  \theta_{a\eta}\right) 
\\
&\Rightarrow~~~ \Gamma(a\to\gamma\gamma)=\frac{g^2_{a\gamma\gamma}}{64\pi} m_a^3. 
\end{align}
The trace runs on the three-flavor space, $Q_q$ is the diagonal matrix with the electric charges of the quarks on the diagonal, $N_c=3$ is the number of colors, $\lambda_3,~\lambda_8$ are Gell-mann matrices (normalization ${\rm tr}[\lambda_a \lambda_b]=\delta_{ab}$), and $\lambda_0=\frac{1}{\sqrt{3}}{\rm diag}\{1,1,1\}$.

\subsubsection{KOTO sensitivity and comparison with other experiments}\label{sec:aGGsensitivity}

\begin{figure*}[t] 
\center
\includegraphics[width=7.2cm]{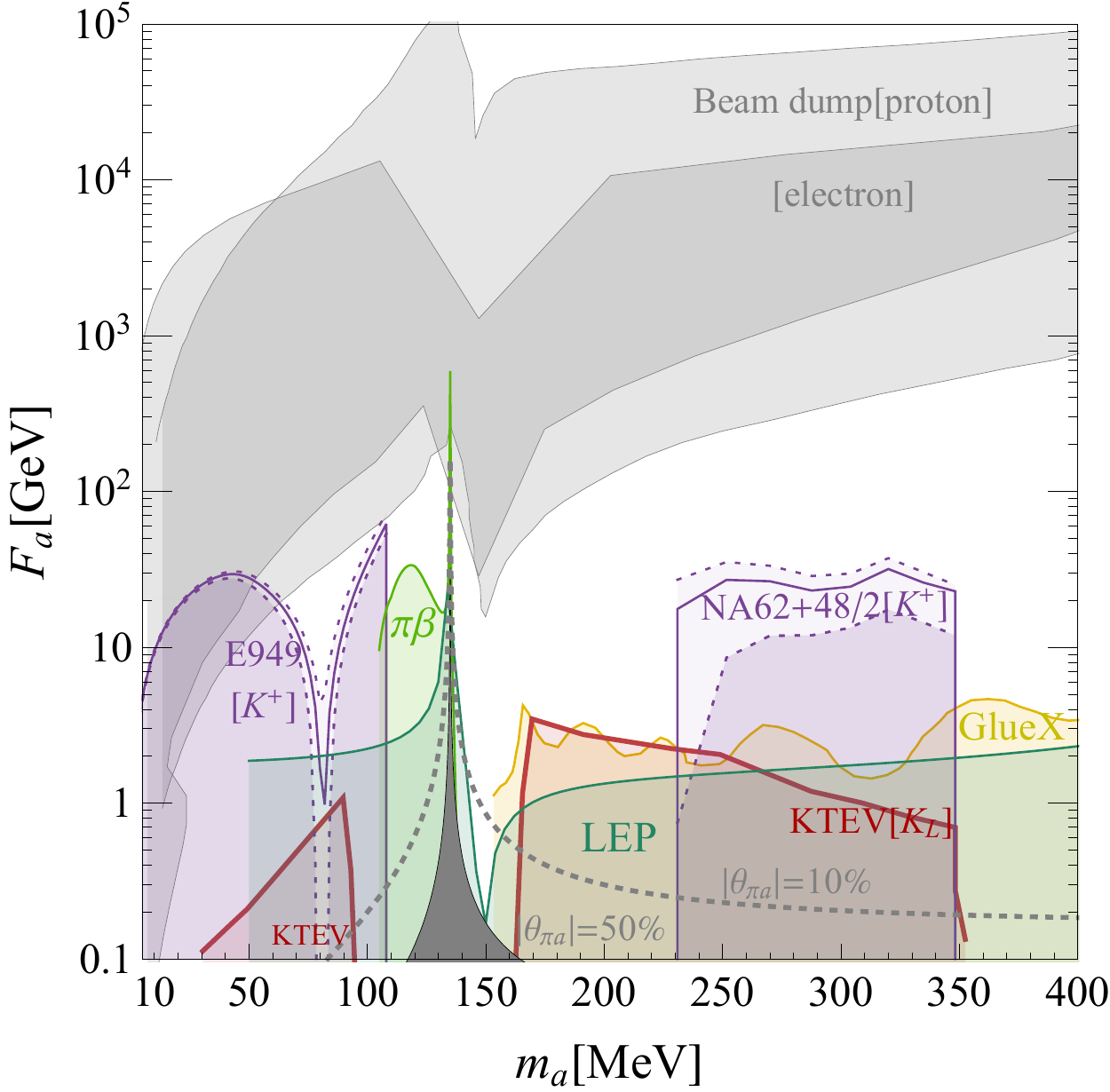}~~~ 
\includegraphics[width=7.2cm]{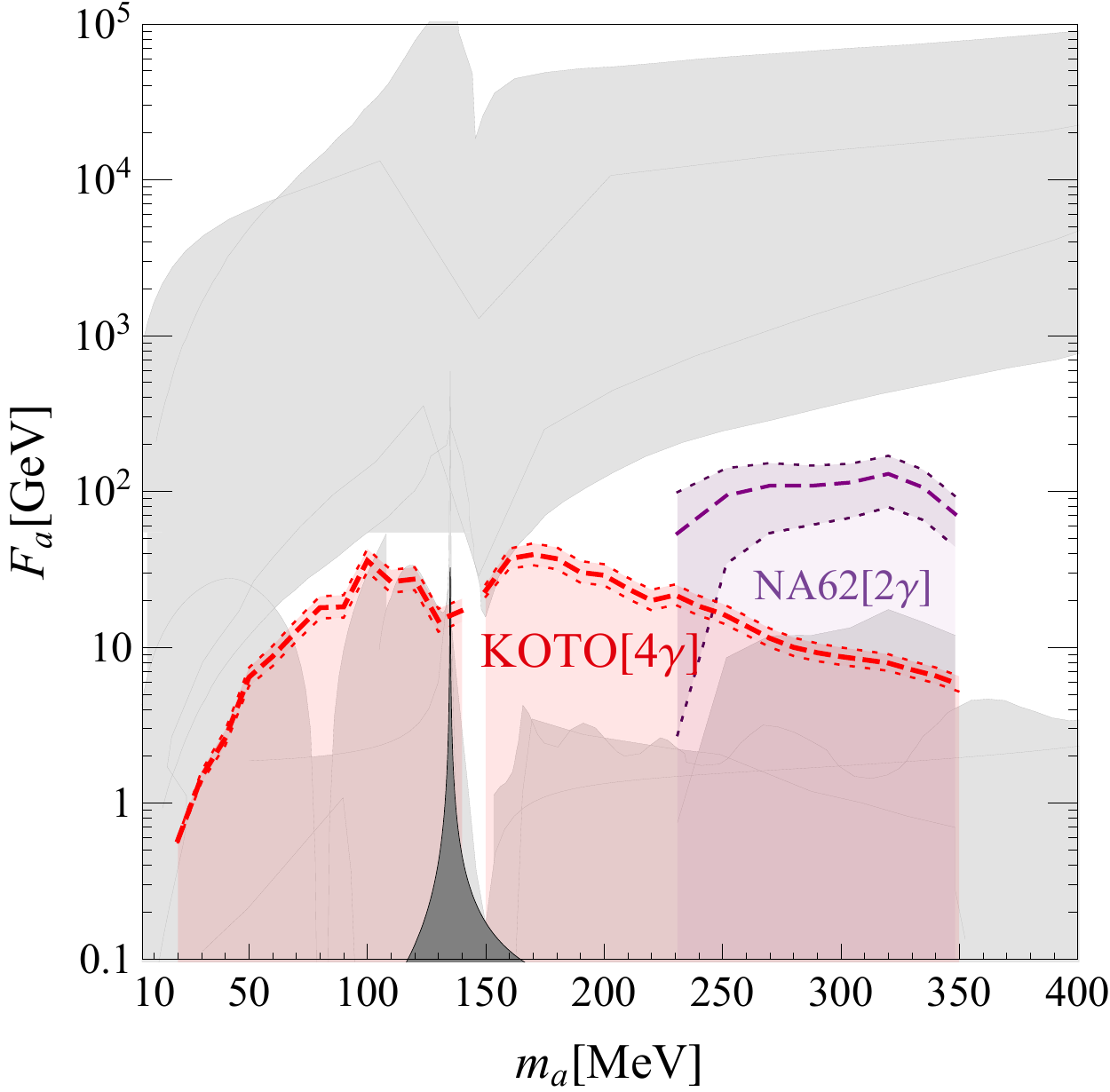} 
\caption{Left panel: Present bounds on the parameter space of the $G\tilde G$ coupled-ALP benchmark, as a function of the ALP mass, $m_a$, and of its decay constant, $F_a$. Right panel: Present and future bounds on the parameter space. In gray, we present the present bound (as shown in the left panel); in red and purple we present the future bounds at KOTO ($4\gamma$ proposed search), and at NA62 ($\pi^++2\gamma$ signature), respectively. 
The bands for the Kaon experiments (E949, NA62, KOTO) show the uncertainties from the quark mass values. See the main text for the discussion. 
}
\label{Fig:BoundGG}
\end{figure*}

In Fig. \ref{Fig:BoundGG}, we show the current bounds (left panel) and future reach (right panel) on the parameter space of this simplified model. We compare the bound from the KOTO experiment to the bounds from other Kaon experiments, as well as 
other present and future accelerator experiments. The discussion for the bounds and projections is almost parallel to Sec.~\ref{sec:aWWsensitivity} for the {\boldmath SU(2)}-coupled ALP simplified model. The most relevant differences arise for  LEP, beam-dump experiments, the GlueX experiment, and PIBETA experiment, which we comment in the following. 
\begin{itemize}
 \item {\bf LEP}\\
The $G\tilde{G}$ coupled ALP does not have a coupling to $Z\gamma$ unlike the {\boldmath SU(2)} coupled ALP.  
Still LEP set a constraint on this benchmark model through the process $e^+e^-\to \gamma^*\to \gamma a$ where the di-photon from the ALP decay is collimated and seen as a single photon. In~\cite{Knapen:2016moh}, the bound on the $aF\tilde{F}$ operator was derived from the OPAL inclusive $2\gamma$ search \cite{Abbiendi:2002je}. We show this bound in dark green in the left panel of Fig.~\ref{Fig:BoundGG}. 

 \item {\bf Proton and electron beam dump experiments}\\
In the proton beam dump experiments, the $G\tilde{G}$ coupled ALP can be produced through the meson mixings and decay by the effective photon coupling. The bound was studied in Ref.~\cite{Ariga:2018uku} using the CHARM result \cite{Bergsma:1985qz}. In our figure, we also include the bound from the electron beam dump experiments, E141 and E137,  where the induced photon coupling is responsible to both the production and decay (see \cite{Dolan:2017osp} and references therein). Both bounds are shown in gray in the left panel of Fig.~\ref{Fig:BoundGG}.

\item {\bf GlueX experiment}\\
The GlueX experiment can be used to set a bound on the ALP parameter space \cite{Aloni:2019ruo}. The experiment utilizes a 9 GeV photon beam colliding against a fixed target.  The ALP can be produced from the decay of vector mesons such as $\rho$ and $\omega$ and observed via its decays to photons. The bound was derived using 1/pb data of \cite{AlGhoul:2017nbp}, and it is shown in yellow in the left panel of Fig.~\ref{Fig:BoundGG}. 

\item {\bf PIBETA experiment}\\
The precision measurement of $\pi^+\to \pi^0(\to\gamma\gamma) e \nu$ at the PIBETA experiment \cite{Pocanic:2003pf} gives a constraint on $\theta_{\pi a}$ for $100~{\rm{MeV}}\lesssim m_a\lesssim m_{\pizero}$ \cite{Altmannshofer:2019yji}. The corresponding bound is shown in light green in the left panel of Fig.~\ref{Fig:BoundGG} (``$\pi\beta$''). We have checked that the measurement of $\pi^+\to e \nu$ by the PIENU collaboration \cite{Aguilar-Arevalo:2015cdf} does not give an additional constraint in the region of parameter space shown in Fig.~\ref{Fig:BoundGG} \cite{Altmannshofer:2019yji}.

\end{itemize}

The main updates in Fig.~\ref{Fig:BoundGG} are the bounds from Kaon decays. In the figure, we only present the bounds obtained from visible searches ($K^\pm\to\pi^+\gamma\gamma$ and $K_L\to\pi^0\gamma\gamma$), since the invisible ones do not extend the reach of the beam dump experiments. We represent each Kaon bound in the figure with a band. This quantifies the uncertainty coming from varying the quark mass ratios in the range $m_u/m_d=0.47~(+0.06,-0.07)$ MeV,  $m_s/\bar m=27.3~(+0.7,-1.3)$ MeV \cite{PDG}. This uncertainty particularly affects the NA62 bound at $m_a\sim 230$ MeV, close to the accidental cancellation for BR$(K^+\to\pi^+ a)$ shown in Fig. \ref{Fig:BRKLGAndLifeTime}. We do not show the uncertainty on the bound coming from the uncertainty in the determination of $\theta_{\eta\eta^\prime}$. This will particularly affect the E949 [$K^+$] bound at low mass. This bound on $F_a$ can change by a factor of $\sim 2$ varying $\theta_{\eta\eta^\prime}$ in $-[10^\circ,19^\circ]$. {Note that the bounds on the $K_L$ decays do not suffer of large uncertainties, as we discuss in more details in Appendix \ref{Sec:AppendixC4}. }

{As shown in the right panel of Fig. \ref{Fig:BoundGG}, in the future, both the NA62 $K^+\to\pi^+a$ (``$[2\gamma]$''), and the proposed KOTO $K_L\to\pi^0 a$ (``$[4\gamma]$'') searches will significantly extend the probed parameter space, especially at $m_a>m_{\pi^0}$.} 
It is also interesting to note that, in the future, the parameter space of the gluon coupled ALP will be fully probed up to decay constants $F_a\sim$ few TeV for $m_a\lesssim m_\pi$. The regions of parameter space not yet probed in the left panel of Fig. \ref{Fig:BoundGG} will be, in fact, probed by the GlueX experiment with 1/fb data \cite{Aloni:2019ruo} and by the SeaQuest experiment at Fermilab \cite{Berlin:2018pwi}.

\section{Breaking the Grossman-Nir bound}\label{sec:GNbreaking}
{In this section, we describe in details the GN-breaking simplified model introduced in Sec. \ref{Sec.2.2}, highlighting the unique sensitivity of the KOTO experiment in probing its parameter space.}
\subsection{Chiral Lagrangian analysis}\label{Sec.ChiralGN}
The effective scalar ($\phi$)-quark couplings given in Eq.~\eqref{EqGN} can be embedded in the chiral Lagrangian in the form of mass terms as
\begin{align}\label{eq:chiralSD}
&\frac{v}{\Lambda_{\rm GNV}^2}\phi^2 \frac{F_\pi^2 B_0 }{\textcolor{blue}{2}} {\rm{Tr}}[y_1 \lambda_{sd} \Sigma ]
+\frac{v}{\Lambda_{\rm GNV}^2}\phi^2 \frac{F_\pi^2 B_0 }{\textcolor{blue}{2}} {\rm{Tr}}[y_2 \lambda_{sd}^\dag \Sigma]+h.c.\,,
\end{align}
where $B_0=m_\pi^2/(m_u+m_d)$, $F_\pi$ is the pion decay constant, $F_\pi\approx93$ MeV, $y_{1,2}$ are real couplings in the mass basis, $v$ is the vacuum expectation value of the Higgs, $v=246$ GeV, and $\Sigma$  is the common exponential pion field matrix 
\begin{align}
    \Sigma\equiv\exp[2i \Pi/F_{\pi}], \quad 
    \Pi\equiv 
   \frac{1}{\sqrt{2}}  \begin{pmatrix}
    \frac{\pi^0}{\sqrt{2}}+\frac{\eta_8}{\sqrt{6}} &  \pi^+ &  K^+ \\
    \pi^- &    -\frac{\pi^0}{\sqrt{2}}+\frac{\eta_8}{\sqrt{6}} & K^0 \\
    K^- & \bar{K}^0 &  -2\frac{\eta_8}{\sqrt{6}}
  \end{pmatrix}\,.
\end{align}
$\lambda_{sd}$ is the three by three matrix leading to $s\to d$ flavor violation
\begin{align}
 \lambda_{sd}=
\begin{pmatrix}
0&0&0\\
0&0&1\\
0&0&0 
\end{pmatrix}\,.
\end{align}
Expanding the Lagrangian in (\ref{eq:chiralSD}) in powers of $\Pi$, we find the matrix elements for the several meson to $\phi$ transitions:
\begin{eqnarray}\label{eq:LagrangiansdExpand}
\mathcal L_{sd}&\simeq& - (y_1+y_2) \frac{v {F_\pi} B_0 }{\Lambda_{\rm GNV}^2} \sigma \chi (K^0+\bar{K}^0)+ i\frac{y_2-y_1}{2} \frac{v {F_\pi} B_0 }{\Lambda_{\rm GNV}^2} (\sigma^2-\chi^2)(K^0-\bar{K}^0)+\\\nonumber
&&+ \frac{y_1+y_2}{2} \frac{v  B_0 }{\Lambda_{\rm GNV}^2}{(\sigma^2-\chi^2)} (K^0+\bar{K}^0)\left(\frac{\pi^0}{\sqrt{2}} +\frac{\eta_8}{\sqrt{6}}  \right)+
\\\nonumber
&&+i (y_2-y_1)\frac{v  B_0 }{\Lambda_{\rm GNV}^2}\sigma\chi (K^0-\bar{K}^0)\left(\frac{\pi^0}{\sqrt{2}} +\frac{\eta_8}{\sqrt{6}}  \right)+\\\nonumber
&& - \frac{y_1+y_2}{2}\frac{v B_0 }{\Lambda_{\rm GNV}^2}{(\sigma^2-\chi^2)} (K^+\pi^-+K^- \pi^+)- i(y_2-y_1)\frac{v B_0 }{\Lambda_{\rm GNV}^2}\sigma\chi (K^+\pi^--K^- \pi^+)\,.
\end{eqnarray}
{These terms will lead to exotic $K^+$, $K_L$ and $K_S$ decays, where we are working on the phase convention $K_L\simeq \frac{K^0+\bar K^0}{\sqrt 2}$ and $K_S\simeq \frac{K^0-\bar K^0}{\sqrt 2}$.} On top of this effective Lagrangian, we add the effective operator in Eq.~\eqref{eq:chiDecay},  $\frac{\chi}{\Lambda_\chi} F_{\mu\nu}\tilde F^{\mu\nu}$,
that is responsible of the $\chi$ decay into two photons. 
As long as $\Lambda_\chi \lesssim {50~{\rm TeV}\left(\frac{m_\chi}{120~\MeV}\right)}^2$, the decay length of $\chi$ {is smaller than $\sim 10$ cm in the analyzed mass range,} and, hence, its decay is effectively prompt~\cite{Kitahara:2019lws}.

\subsection{New Kaon decays}\label{Sec:NewDecaysGNModel}
From the effective Lagrangian in (\ref{eq:LagrangiansdExpand}), we can compute the matrix elements for the several transitions. We find

\begin{align}
\Gamma(K_L\to\sigma\chi)=\frac{1}{4\pi}\left(y_1+y_2\right)^2 \frac{v^2 F_\pi^2 B_0^2}{\Lambda_{\rm GNV}^4}\frac{|\vec{p}_\sigma|}{m_{K_L}^2}\,,
\end{align}
where $|\vec{p}_\sigma|$ is the absolute value of the momentum of $\sigma$ in the center of mass frame. 

Also the charged Kaons will inherit new exotic decay modes. However, due to charge conservation, only decay modes with three (or more) final states will be generated (see (\ref{eq:LagrangiansdExpand})). In particular,
\begin{eqnarray}\label{eq:KpmToPisigmasigma}
\Gamma(K^\pm\to\pi^\pm\sigma\sigma)&=&\frac{1}{128\pi^3}\left(\frac{(y_1+y_2)^2}{4}\frac{v^2B_0^2}{\Lambda_{\rm GNV}^4}\right)\frac{\int{dm_{\sigma\sigma}^2dm_{\sigma\pi}^2}}{m_K^3},\\
\Gamma(K^\pm\to\pi^\pm\sigma\chi)&=&\frac{1}{256\pi^3}\left((y_1-y_2)^2\frac{v^2B_0^2}{\Lambda_{\rm GNV}^4}\right)\frac{\int{dm_{\sigma\chi}^2dm_{\sigma\pi}^2}}{m_K^3}.
\end{eqnarray}
Analogously, $K^\pm$ can also decay to $\pi^\pm\chi\chi$ with the amplitude given by (\ref{eq:KpmToPisigmasigma}) with the replacement $dm_{\sigma\sigma}^2dm_{\sigma\pi}^2\to dm_{\chi\chi}^2dm_{\chi\pi}^2$.

$K_L$ also acquires new three-body decays:
\begin{equation}\label{eq:KLThree}
\Gamma(K_L\to\pi^0\sigma\sigma)=\frac{1}{128\pi^3}\left(\frac{(y_1+y_2)^2}{4}\frac{v^2B_0^2}{\Lambda_{\rm GNV}^4}\right)\frac{\int{dm_{\sigma\sigma}^2dm_{\sigma\pi}^2}}{m_{K_L}^3}\,,
\end{equation}
and correspondingly for $K_L\to\pi^0\chi\chi$.

Finally, new (two or three-body) decays of $K_S$ will be also induced:
\begin{eqnarray}\label{eq:KS}
\Gamma(K_S\to\sigma\sigma)&=&\frac{1}{8\pi}\left(y_1-y_2\right)^2 \frac{v^2 F_\pi^2 B_0^2}{\Lambda_{\rm GNV}^4}\frac{|\vec{p}_\sigma|}{m_{K_S}^2}\,,\\
\Gamma(K_S\to\pi^0\sigma\chi)&=&\frac{1}{256\pi^3}\left((y_1-y_2)^2\frac{v^2B_0^2}{\Lambda_{\rm GNV}^4}\right)\frac{\int{dm_{\sigma\chi}^2dm_{\sigma\pi}^2}}{m_{K_S}^3}\,,
\end{eqnarray}
and, similarly, one can obtain the width for $K_S\to\chi\chi$ with the replacement $|\vec{p}_\sigma|\to|\vec{p}_\chi|$. {These decay modes will be obviously more suppressed due to the larger width of $K_S$.}

\subsection{KOTO sensitivity and comparison with other Kaon measurements}
The decay mode $K_L\to\sigma\chi$ can show up in the KOTO $K_L\to\pi^0\nu\bar\nu$ signal region, as long as the pseudoscalar $\chi$ has a short enough life-time, {and $\sigma$ is stable in the KOTO detector}. In particular, one needs $\tau_\chi\lesssim 10$ cm that implies the operator $\frac{\chi}{\Lambda_\chi} F_{\mu\nu}\tilde F^{\mu\nu}$ to be suppressed by a not too large NP scale: $\frac{1}{\Lambda_\chi}\gtrsim 1/{50}$ TeV$^{-1}$.

This term induces kinetic mixing between $\chi $ and the pion at one loop,
$\epsilon\,  \partial_\mu \chi \partial^\mu \pi^0$, with 
$\epsilon \sim 1/16\pi^2 \times g_{\pi\gamma} \, g_{\chi\gamma}\, \Lambda_{\rm cutoff}^2$ 
with $g_{\pi\gamma}$ and $g_{\chi\gamma}$ being the pion and $\chi$ photon-couplings, respectively ($g_{\pi\gamma} = \alpha/ (4\pi F_\pi)$ and $g_{\chi\gamma}=1/\Lambda_\chi$), leading \beq
\epsilon \sim 10^{-9} \times \left({ \Lambda_{\rm cutoff}\over\rm GeV}\right)^2\times{ {50}\rm\,TeV\over\Lambda_\chi}\,.
\eeq
The loop is quadratically sensitive to the internal momentum, $\Lambda_{\rm cutoff}$. The loop momenta that characterize the pion-photons coupling decrease significantly above the QCD scale.
Therefore, the above estimate of the $\chi-\pi^0$ mixing shows that this effect can be neglected. 
As for $\sigma$, it can decay to four photons (e.g. via its coupling to $\chi$ and a neutral Kaon which couples to two photons) however this coupling is suppressed by CKM factors, extra loop and $1/\Lambda_\chi$. Therefore it is safe to consider $\sigma$ effectively stable.

 \begin{figure*}[t!]
 \begin{center}\label{KOTOSignal}
  \includegraphics[width=0.45\linewidth]{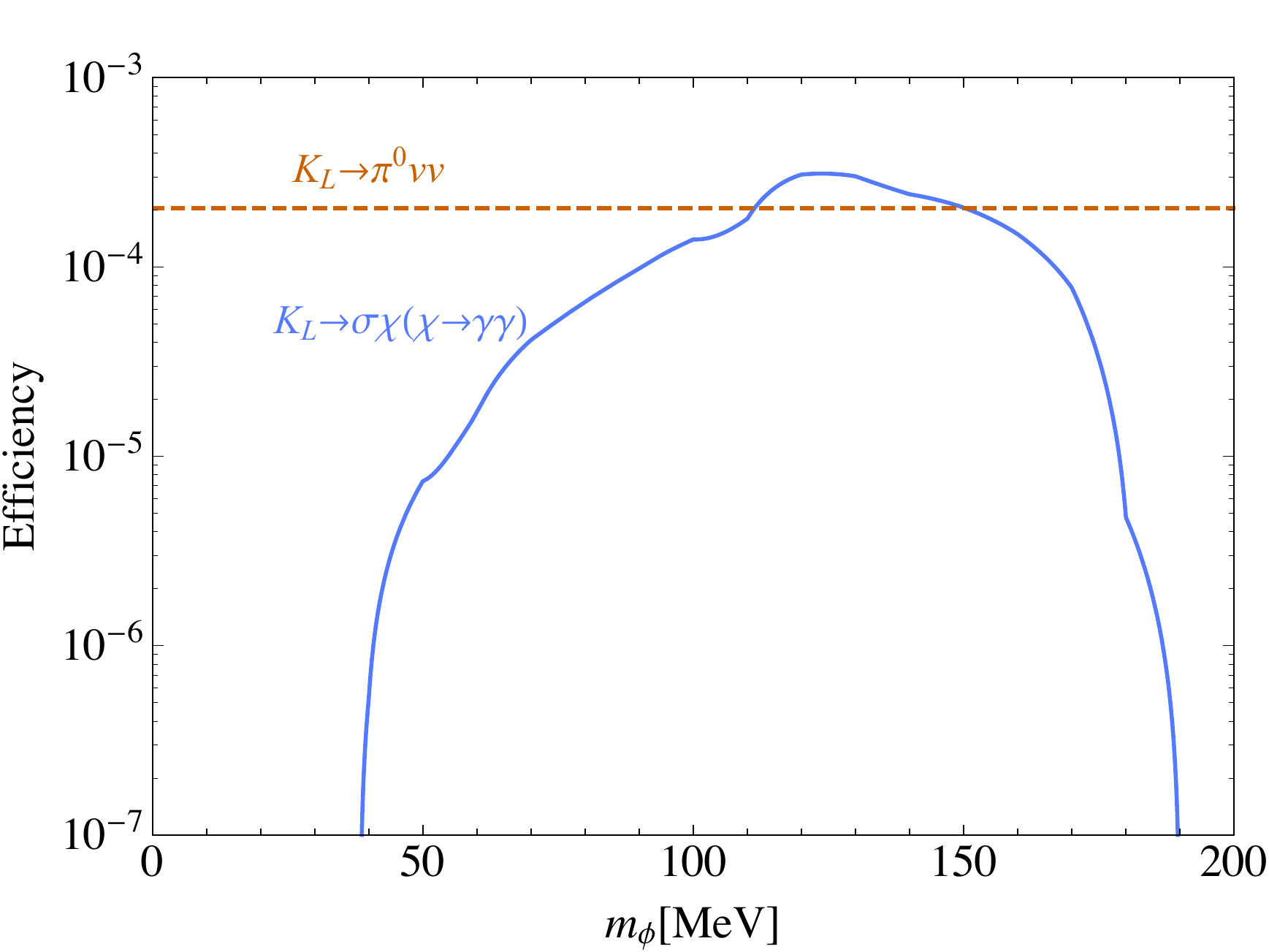}~~~~~~
    \includegraphics[width=0.45\linewidth]{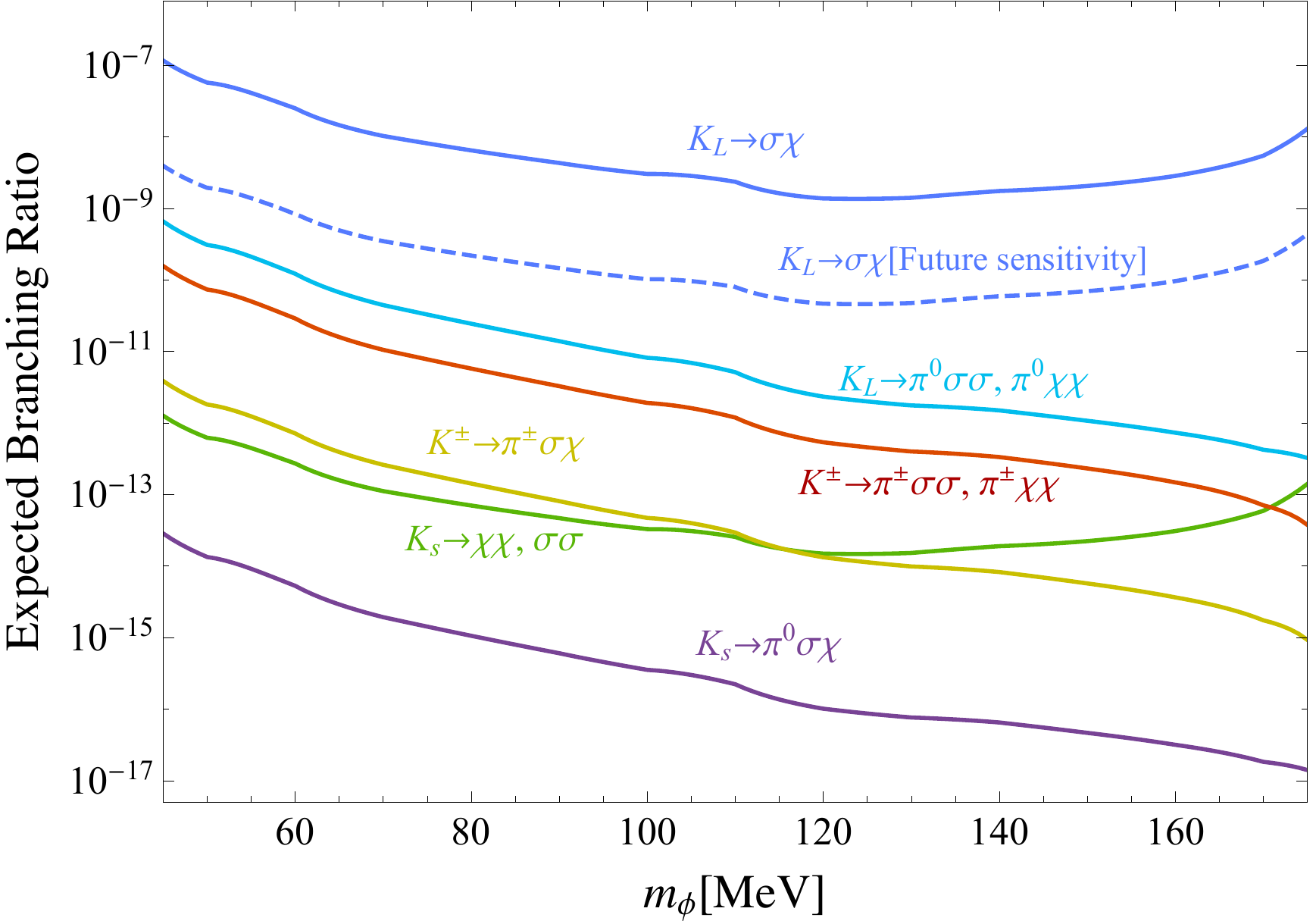}
 \end{center}
 \caption{Left panel: signal efficiency as a function of the $\chi$ mass (blue curve). In the plot, we fix $m_\chi=m_\sigma$. For comparison, we show in red the KOTO efficiency for the $K_L\to\pi^0\nu\bar\nu$ signal. Right panel: {the blue lines represent the BR$(K_L\to\sigma\chi)$ needed to produce 3 events in the KOTO signal region using the data collected in 2016-2018 (solid line), or future KOTO data (dashed blue).} The other curves correspond to the predictions for BR$(K_L\to\pi^0\sigma\sigma)$ and BR$(K_L\to\pi^0\chi\chi)$ (light blue), BR$(K^\pm\to\pi^\pm\sigma\sigma)$ and BR$(K^\pm\to\pi^\pm\chi\chi)$(red), BR$(K^\pm\to\pi^\pm\sigma\chi)$ (yellow), BR$(K_S\to\sigma\sigma)$ and BR$(K_S\to\chi\chi)$ (green), and BR$(K_S\to\pi^0\sigma\chi)$ (purple), {once we demand the model to produce 3 events in the KOTO signal region using the data collected in 2016-2018. For the latter three curves, we have fixed $y_2=2y_1$.}
 }
\end{figure*}

 \begin{figure*}[t!]
 \begin{center}\label{KOTOdecayvol}
  \includegraphics[width=0.4\linewidth]{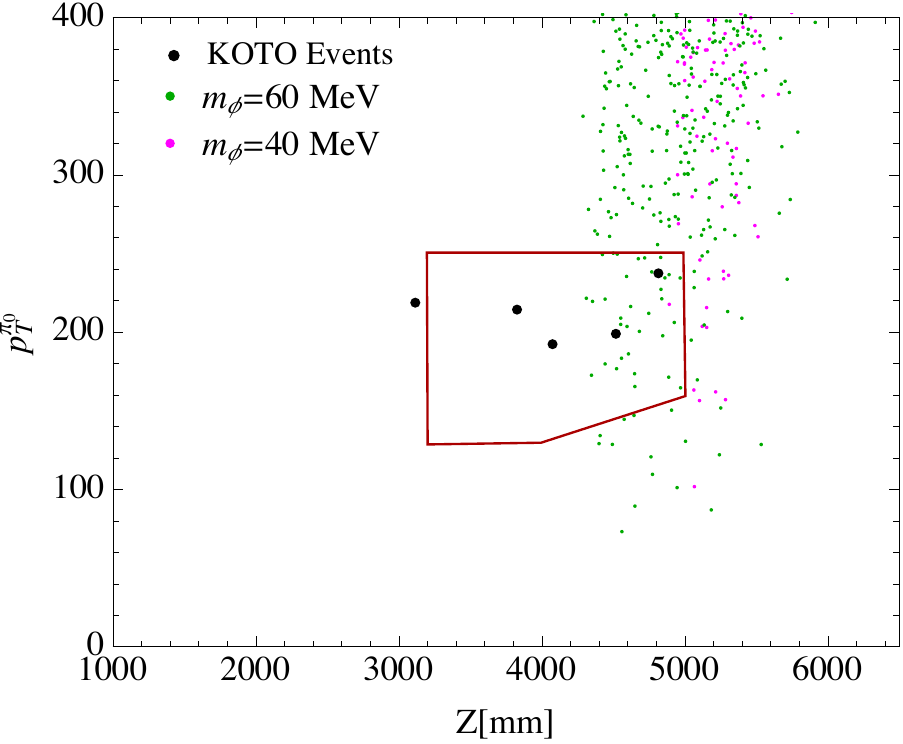}
  \hspace{5pt}
  \includegraphics[width=0.4\linewidth]{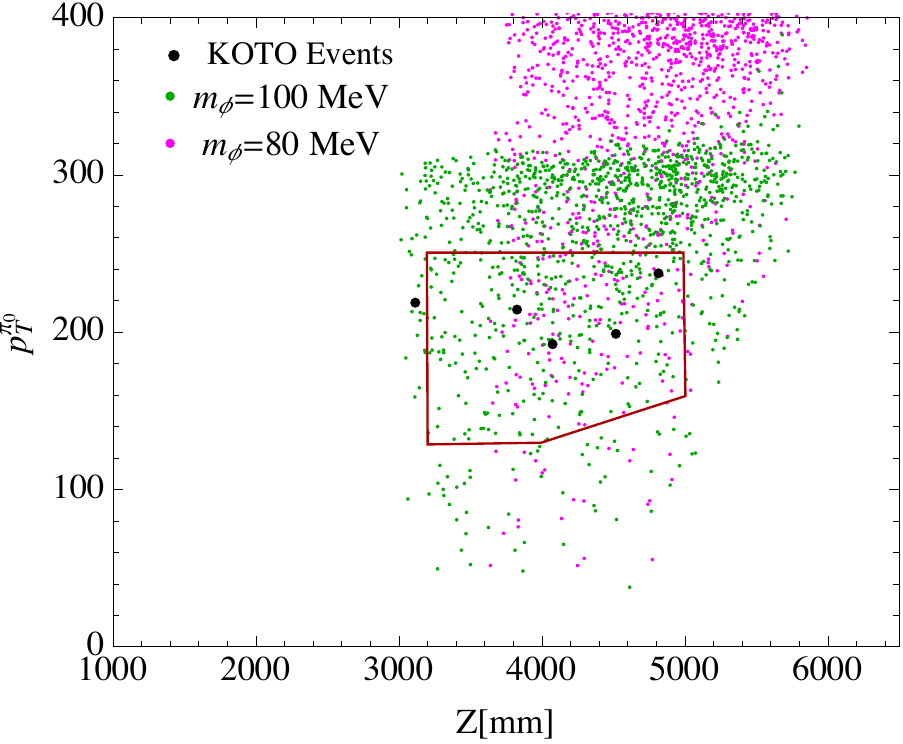}
  \includegraphics[width=0.4\linewidth]{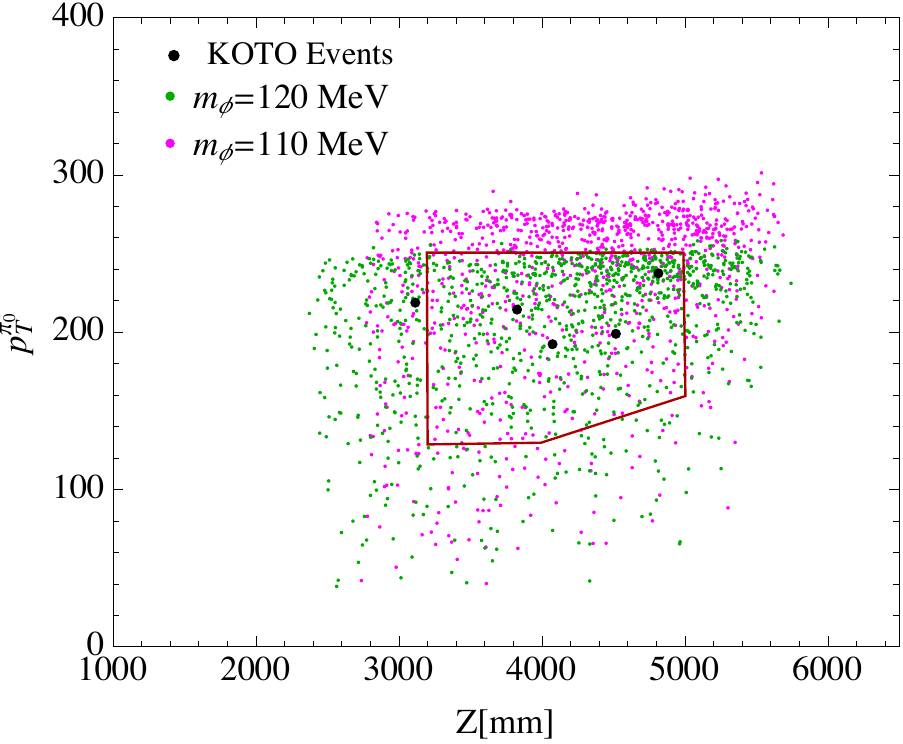}
    \hspace{5pt}
  \includegraphics[width=0.4\linewidth]{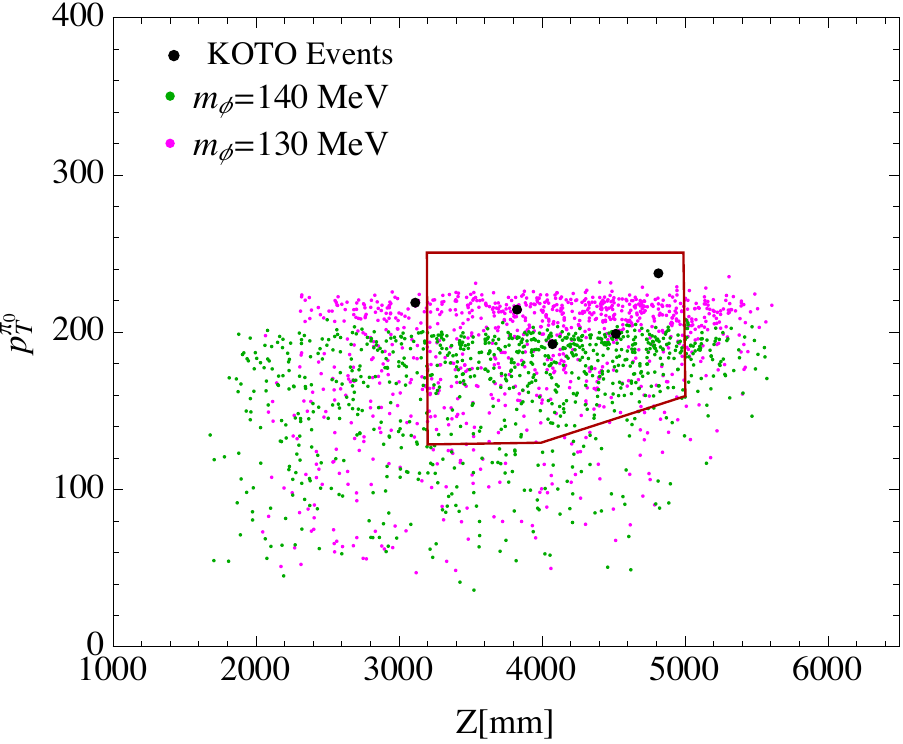}
  \includegraphics[width=0.4\linewidth]{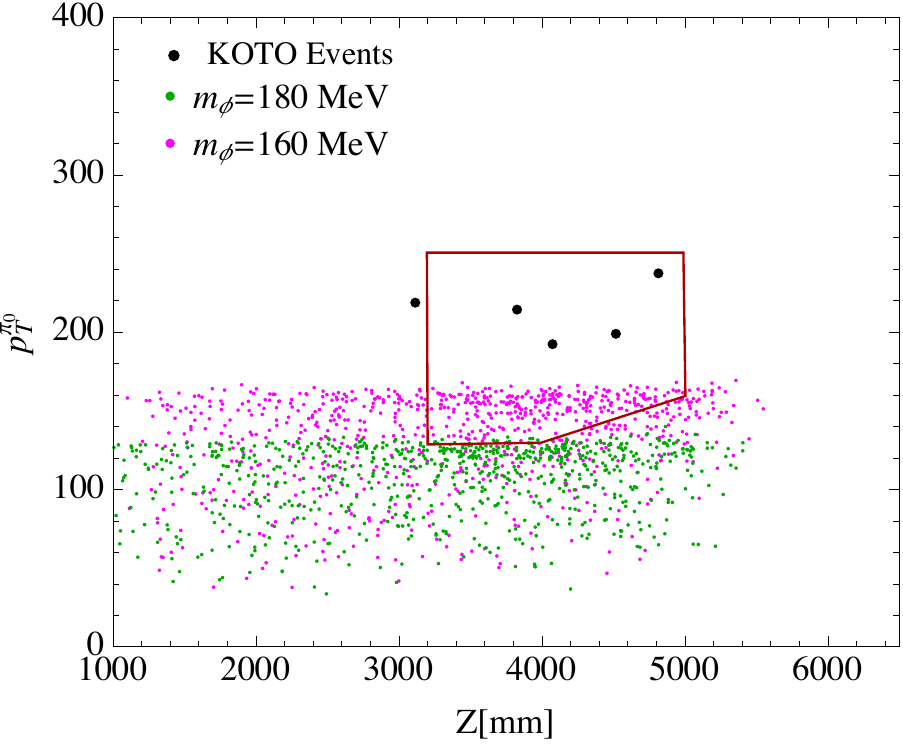}
    \hspace{5pt}
  \includegraphics[width=0.4\linewidth]{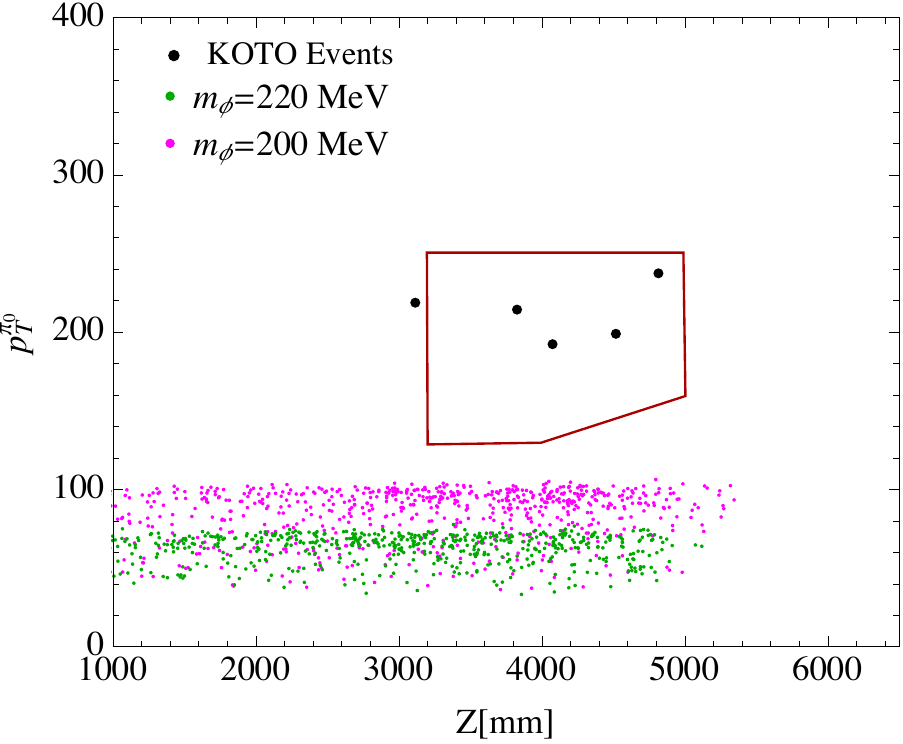}
 \end{center}
 \caption{The expected event distribution of $K_L\to \sigma\chi$ decays for different mass points $m_\phi\equiv m_\sigma= m_\chi$. We simulate $10^4$ decays for each mass point, reconstruct the vertex and momentum, and apply the kinematic cuts of the KOTO $K_L\to \pi^0\nu\bar\nu$ analysis. {The KOTO signal region is delimited in red, as a function of the position of the vertex, $Z$, and of the reconstructed di-photon transverse momentum, $p_T^{\pi 0}$.}}
\end{figure*}

The efficiency for $K_L\to\sigma\chi$ to end up in the KOTO signal region depends crucially on the mass of the $\sigma$ and $\chi$ particles. In the left panel of Fig. \ref{KOTOSignal}, we show in blue the efficiency has a function of $m_\chi$, that, for convenience, we fix to be $= m_\sigma$. A sizable efficiency is reached as long as the $\chi$ mass is not too far away from the mass of the pion. In Fig. \ref{KOTOdecayvol}, we also show the distribution of our montecarlo events for $K_L\to\sigma\chi$ for different values of the $m_\chi=m_\sigma$ mass. As we can observe, the events fall nicely in the signal region (the region delimited in red) as long as $100~{\rm{MeV}}\lesssim m_\chi=m_\sigma \lesssim 160$ MeV. 

Using the efficiency of the left panel of Fig.~\ref{KOTOSignal} and the widths discussed in the previous section, we can compute the sensitivity of KOTO to our model, as well as the corresponding predictions for the other exotic $K^+$ and $K_S$ decay modes. In the right panel of Fig. \ref{KOTOSignal}, {the blue lines represent the BR$(K_L\to\sigma\chi)$ needed to produce 3 events in the KOTO signal region using the data collected in 2016-2018 (solid line), or future KOTO data (dashed blue).} Note that 2016-2018 data is already able to probe {a branching ratio as small as BR$(K_L\to\sigma\chi)\sim 1.3\times 10^{-9}$. This corresponds to a GN breaking scale as high as $\Lambda_{\rm GNV}/\sqrt{(y_1+y_2)}\sim 10^7$ GeV.}

The other lines in the right panel of Fig.~\ref{KOTOSignal} are the corresponding predictions for BR$(K_L\to\pi^0\sigma\sigma)$ and BR$(K_L\to\pi^0\chi\chi)$ (light blue), BR$(K^\pm\to\pi^\pm\sigma\sigma)$ and BR$(K^\pm\to\pi^\pm\chi\chi)$(red), BR$(K^\pm\to\pi^\pm\sigma\chi)$ (yellow), BR$(K_S\to\sigma\sigma)$ and BR$(K_S\to\chi\chi)$ (green), and BR$(K_S\to\pi^0\sigma\chi)$ (purple), {once we demand the model to produce 3 events in the KOTO signal region using the data collected in 2016-2018.} For the latter three curves, we have fixed $y_2=2y_1$ {(see the parametric dependence of the several widths discussed in Sec. \ref{Sec:NewDecaysGNModel}).}

\section{Discussion and overview}\label{Sec:summary}
Rare Kaon decay modes have been always considered among the few holly grails of flavor physics because of their rareness, and because of our ability to control them well theoretically.
This made rare kaon decays singular in their ability to probe new physics (NP) models. 

The current time is rather unique as both the KOTO and the NA62 experiments are collecting high quality data aiming to reach unprecedented precision in the measurement of neutral and charged Kaons, respectively, providing a direct test to the SM predictions. What makes all this possible is the huge fluxes of Kaons achieved at J-PARC and at CERN. 
The fact that the KOTO and NA62 detectors have access to these astronomical fluxes makes them sensitive to other types of dynamics that we denote as ``dark sector physics''. 
By dark sectors we refer to a class of models with light particles that couple only weakly to the Standard Model (SM) fields. In this work, we have shown that such dark sectors can be probed in regions that could not have been searched for so far, which is rather exciting. 

In this paper, we particularly highlight the complementarity of the two experiments. Naively, the fact that NA62 already probes charge Kaon decays with branching ratios as small as} $10^{-10}$, while KOTO is an order of magnitude behind in the corresponding neutral decay mode, makes one conclude that KOTO is only providing us with a secondary validation of the searches done at the NA62. This statement may be enforced by the Grossman-Nir (GN) relation that bounds the size of the NP contributions in the neutral mode via the charged one.
We, however, demonstrate that the physics of dark sectors do not necessarily follow this pattern.
The reason is two fold: i) on the experimental side, the two experiments are different in several essential aspects, in terms of kinematics, acceptance and sensitivity to different final states;
ii) on the theoretical side, when examining dark sectors one find that the NP-GN relation can be effectively violated by as much as several orders of magnitude.

We show this by considering two qualitatively different physics cases. 
{\it{In the first}}, we consider models of axion-like-particles (ALPs) which couple to electroweak gauge bosons or to gluons. We find that ALP-diphoton decay mode, $K_L\to \pi^0(\gamma\gamma) a(\gamma\gamma)$, can be very efficiently searched for at KOTO.
We layout a new search strategy that, if adopted by the collaboration, would allow KOTO to probe uncharted territories of ALP-physics. At the same time we also find that the corresponding final state at NA62, $K^+\to\pi^+ a$, while being equally interesting, can suppressed. This is the case of the ALP-coupled to gluon model where cancellations between different contributions when including the $\eta$ and $\eta'$ contributions can happen, albeit with large theoretical uncertainties. This probably calls for a more detailed theoretical analysis of the $K^+\to\pi^+ a$ decay, going beyond leading order in the chiral Lagrangian, and also carefully including the uncertainties related to quark masses and to the $\eta-\eta'$ mixing. This might be an interesting study to be performed on the lattice which would then be freed from the uncertainties related to the chiral expansion. 
{\it{In the second}}, we introduce a model based on approximate strange flavor symmetry, that effectively leads to a strong violation of the Grossman-Nir bound.
We find that this benchmark model can be discovered by the KOTO experiment looking for $K_L$ decaying into two photons plus invisible. It is also worthwhile to mention that this benchmark could also account for the potential-candidate events seen at KOTO and the absence of signals at NA62 at the same time. In fact, the corresponding charged Kaon signals at NA62 would be several orders of magnitude suppressed, and effectively hidden to this latter experiment.

Our main messages of this paper are quite general and motivate model independent searches at both Kaon factories.

\bigskip
\underline{Note added:} while this work was at its final stage of completion, Refs.~\cite{Ziegler:2020ize,He:2020jly,Liao:2020boe, Descotes-Genon:2020buf} that have some overlap with with the topics discussed above, appeared.
\section*{Acknowledgements}

We thank Wolfgang Altmannshofer, Gaia Lanfranchi, and Hajime Nanjo for the useful discussion.  The research of SG is supported in part by the NSF CAREER grant PHY-1915852. SG would like to thank the Aspen Center for Physics under NSF grant PHY-1607611, where part of this work was performed. 
The work of GP is supported by grants from The U.S.-Israel Binational Science Foundation (BSF), European Research Council (ERC), Israel Science Foundation (ISF), Yeda-Sela-SABRA-WRC, and the Segre Research Award. KT is supported by the US Department of Energy grant DE-SC0010102. 

\appendix 
\section{SM reconstruction} \label{sec:SMrecon}
The  axion reconstruction we introduce in Sec. \ref{sec:recon} should also work for the SM processes $K_L\to \pi^0\pi^0, \pi^0\gamma\gamma$. A dedicated algorithm was employed for $K_L\to \pi^0\pi^0$ by the KOTO collaboration. 
This algorithm is well-tested and, indeed, it has been used for flux measurements of the incoming $K_L$.  We use this method to cross check our MC simulation.  We briefly review it in the following. 

The strategy is:
\begin{enumerate}
\vspace{-0.3cm}
\item
Assume the four photons come from two neutral pions, and assign the four photons to two pairs, say $\gamma_1,\gamma_2$ and $\gamma_3,\gamma_4$. There are three possible combinations. 
\item The position of the vertex, $Z_{{\rm vtx}}$, is reconstructed based on each pair of photons from the requirement
\begin{align}
m_{\gamma_1\gamma_2}^2(Z_{\rm vtx,1})\equiv m_{\pi^0}^2, \quad
m_{\gamma_3\gamma_4}^2(Z_{\rm vtx,2})\equiv m_{\pi^0}^2. 
\end{align}
Each of this equation leads to at most two solutions for $Z_{\rm vtx,1}$ and $Z_{\rm vtx,2}$.

\item
Pick the combination where the two reconstructed vertices, $Z_{\rm vtx,1}$ and $Z_{\rm vtx,2}$, are the most consistent. A ``pairing variance'' is introduced to evaluate the consistency of the two vertices,  
\begin{align}
&\chi_{dz^2}^2=\sum_{i=1}^{n_{\rm pair}} \frac{(dz^2_i-\overline{dz^2})^2}{\sigma^2_{dz^2,i}} , 
\\
&\overline{dz^2}\equiv \left(\sum_{i=1}^{n_{\rm pair}}\frac{dz_i^2}{\sigma^2_{dz^2,i}}\right)\bigg/\sum_{i=1}^{n_{\rm pair}}\frac{1}{\sigma^2_{dz^2,i}}\,,
\end{align}
where $n_{\rm pair}=2$  for four photon events, and $dz$ is the {distance} from the ECAL ($Z_{\rm vtx}=6.148{\rm m}-dz$). {The combination that minimizes} $\chi_{dz^2}^2$ is picked, and then the decay vertex of $K_L$ is identified. Since $dz^2=dz^2(\vec{r}_1,\vec{r}_2, E_{\gamma_1},  E_{\gamma_2})$,  the variance is obtained by the combination of resolutions of photon position and energy:
\begin{align}
\sigma^2_{dz^2,1}&=\left(\frac{\partial dz^2_i}{\partial r_{\gamma_1}} \sigma_{r_{\gamma_1}} \right)^2 \oplus\left(\frac{\partial dz^2_i}{\partial r_{\gamma_2}} \sigma_{r_{\gamma_2}}\right)^2 \oplus
\left(\frac{\partial dz^2_i}{\partial E_{\gamma_1}} \sigma_{E_{\gamma_2}} \right)^2 \oplus\left(\frac{\partial dz^2_i}{\partial E_{\gamma_2}} \sigma_{E_{\gamma_1}}\right)^2 
\\
&=\left(\frac{\partial dz^2_i}{\partial r_{\gamma_1}} \sigma_{r_{\gamma_1}} \right)^2\oplus\left(\frac{\partial dz^2_i}{\partial r_{\gamma_2}}\sigma_{r_{\gamma_2}} \right)^2 
\nonumber\\&\oplus
\left(\frac{\partial dz^2_i}{\partial \cos\theta}\frac{1-\cos\theta}{E_{\gamma_1}} \sigma_{E_{\gamma_1}} \right)^2 
\oplus
\left(\frac{\partial dz^2_i}{\partial \cos\theta}\frac{1-\cos\theta}{E_{\gamma_2}} \sigma_{E_{\gamma_2}} \right)^2 \,,
\end{align}
where $\sigma_E$ and $\sigma_r$ are the energy and position resolution, respectively. 
\end{enumerate}

\section{Validation of our analysis} \label{sec:validation}
We validate our simulation and reconstruction algorithm by cross-checking the measured quantities at KOTO \cite{Masuda:2015eta, Ahn:2018mvc}.  

\subsection{Detector effects}
First, we check the the detector smearing we include in our analysis reproduces the KOTO results. At KOTO, $\pizero \to \gamma\gamma$ was measured in a special run and the shape of the di-photon invariant mass  was reported in Fig.~7 of \cite{Masuda:2015eta}. 
Assuming the decay vertex is known, we reproduce the shape with the energy and position resolution. We use the detector parameters given in \cite{Masuda:2015eta}. We find that the position resolution is only a minor effect.  This implies that, when the vertex is reconstructed, the main source of uncertainties is from ECAL smearing.

\subsection{Reconstruction of four photons}
$K_L\to 4\gamma$ was measured by the KOTO collaboration, and the reconstructed four photon invariant mass is shown in Fig. 11 of \cite{Masuda:2015eta}, . The peak region is dominated by $K_L\to \pizero\pizero$. 
We simulate $K_L\to \pizero\pizero$ events and perform the SM reconstruction discussed in Appendix~\ref{sec:SMrecon}. The shape of the peak region is well reproduced.

Ref. \cite{Masuda:2015eta} provides the acceptance for the performed analysis, $A=1.48\times 10^{-3}$. 
We reproduce this acceptance at the 10\% level, as it is shown in Table~\ref{cutflow}. 

\begin{table}
\begin{center}
  \begin{tabular}{ |c | r | r | }
    \hline
    Process & Cut Flow & Acceptance \\
    \hline
    \hline
   $K_L$@Beam Exit&  & 1 \\
    \hline
   $K_L$ decay in $2<Z<6.148m$ (truth level)& 100,000 &  7.86\%\\
    \hline
   4 photons hit ECAL &  5,085&  \\
    \cline{1-2}
   $R_{max}\leq$85cm&  3,792 &  \\
    \cline{1-2}
   $\min{E_{\gamma_i}}\geq$ 50 MeV&  3,442&  \\
    \cline{1-2}
   $d_{min}\geq$15cm&  3,056&  \\
    \cline{1-2}
   $\Delta m_{\pi}\leq$ 6 MeV&  2,650&  \\
    \cline{1-2}
   $\Delta m_{K_L}\leq$ 15 MeV&  2,473&  \\
    \cline{1-2}
   $2m\leq Z_{\rm vtx}\leq 5.4m$&  2,115&  \\
    \cline{1-2}
   $\sum E_{1/2}$&  2,011&  \\
    \hline
   $|X_{coe}|\leq$ 6 cm, $|Y_{coe}|\leq$ 6 cm&  2,011& 1.58$\times 10^{-3}$ \\
    \hline
   \end{tabular} 
  \end{center} 
  \caption{Decay probability, cut flows, and acceptance of our $K_L\to \pi^0\pi^0$ analysis. }\label{cutflow}
\end{table}

\subsection{Analysis of the decay \boldmath $K_L\to \pizero\nu\bar\nu$}
We have used the KOTO $K_L\to \pizero\nu\bar\nu$ result {to set constraints on the two ALP benchmarks of Sec. \ref{sec:simplified}}, as well as for the $K_L\to \sigma\chi$ analysis. 
First, we cross checked the efficiency of $K_L\to \pizero X(\to invisible)$ with \cite{Kitahara:2019lws}. We calculated the KOTO correction factor $\epsilon(K_L\to \pizero a)$.

Also, we checked the overall acceptance. With the efficiencies of veto and shower cut ($\epsilon_{\rm veto}=17\%$ and $\epsilon_{\rm shower}=52\%$), we get an acceptance $\sim 30\%$ higher than the reported acceptance reported in \cite{Ahn:2018mvc}. This information is useful to {estimate the uncertainty} of our proposed $K_L\to \pizero a\to 4\gamma$ search, where the result is fully based on our simulations.  

\section{\boldmath$ \Delta S=1$ transitions and ALP coupled to gluons}\label{Appendix:ALPglue}

\subsection{ALP-meson mixing}
We want to compute the ALP-meson mixing arising from the the effective Lagrangian in \eqref{eq:eftGG}. We work in the framework of the chiral Lagrangian and perform a chiral rotation such to remove the mass-mixing between the ALP and the light mesons \cite{Georgi:1986df}. 
The remaining ALP interactions with SM mesons is through the kinetic mixing that is given by
\begin{align}
        &{{\cal L}_{eff}= \frac{iF_\pi^2}{4}\frac{\partial_\mu a}{F_a}
    {\rm Tr}[\tilde\kappa_q(\Sigma^\dag{D}^\mu \Sigma-\Sigma{D}^\mu \Sigma^\dag)] }\,,
    \label{eq:ALPkinmixing}
\end{align}
where $\tilde\kappa_q$ is the diagonal matrix with $\kappa_q =\frac{1}{m_q}/\sum_{q'}(\frac{1}{m_{q'}})$ on the diagonal, and the $\Sigma$ is the non-linear meson field. However, due to the non-negligible mixing between the $\eta$ and $\eta'$ mesons, the $\eta'$ meson has a mass mixing with the ALP through the axial anomaly. 

To compute this effect, we follow the prescription given in \cite{Aloni:2018vki}. We keep only the light state $\eta$ in the mass basis and decouple the $\eta^\prime$, assuming $\sin{\theta_{\eta\eta'}}=-1/3$ (see, however, Eq. (\ref{eq:ThetaALPEta}) for the generic expression of the ALP-$\eta$ mixing). Then the non-linear meson field is given by
\begin{align}\label{eq:SigmaChiral}
  {  \Sigma\equiv\exp[2i \Pi/F_\pi], \quad 
    \Pi\equiv 
    \frac{1}{\sqrt{2}} \begin{pmatrix}
    \frac{\pi^0}{\sqrt{2}}+\frac{\eta}{\sqrt{3}} &  \pi^+ &  K^+ \\
    \pi^- &    -\frac{\pi^0}{\sqrt{2}}+\frac{\eta}{\sqrt{3}} & K^0 \\
    K^- & \bar{K}^0 & -\frac{\eta}{\sqrt{3}},
  \end{pmatrix}}
\end{align}
where we are adopting the pion decay constant $F_\pi \sim 93~\MeV$. The chiral Lagrangian has now both kinetic mixing terms of the ALP with the pion and the $\eta$, and mass mixing terms of the ALP with the $\eta$:
\begin{align}
        &{{\cal L}_{eff}= \frac{iF_\pi^2}{4}\frac{\partial_\mu a}{F_a}
    {\rm Tr}[\tilde\kappa_q(\Sigma^\dag{D}^\mu \Sigma-\Sigma{D}^\mu \Sigma^\dag)] }+{\frac{F_\pi^2}{2}B_0 {\rm Tr}[\Sigma m^\dagger+m^\dagger\Sigma^\dagger]}\,,
    \label{eq:ALPkinAndMassmixing}
\end{align}
{where $m$ is the matrix $m={\rm{exp}}(i\kappa_q\frac{a}{2F_a}\gamma_5)m_q {\rm{exp}}(i\kappa_q\frac{a}{2F_a}\gamma_5)$} and $B_0=m_\pi^2/(m_u+m_d)$. After diagonalizing this system, the physical ALP and meson eigenstates are given by 
    \begin{align}
    \begin{pmatrix}
a\\
\pi\\
\eta
\end{pmatrix}
\simeq
\begin{pmatrix}
       1 & -\frac{\Kpi \Mpi^2}{\Mpi^2-\Ma^2} 
        &  -\frac{\Keta \Meta^2+\delta M_{\eta a}}{\Meta^2-\Ma^2}  \\
         \frac{\Kpi \Ma^2}{\Mpi^2-\Ma^2} &1 & 
          0 \\
        \frac{\Keta \Ma^2+\delta M_{\eta a}}{\Meta^2-\Ma^2}         & 0 & 1
\end{pmatrix}
\begin{pmatrix}
a_{\rm phys}\\
\pi_{\rm phys}\\
\eta_{\rm phys}
\end{pmatrix},
\end{align}
where
\begin{align}
   &
   \Kpi=-\frac{F_\pi}{2F_a}(\kappa_u-\kappa_d), \quad
         \Keta=-\frac{F_\pi}{\sqrt{6}F_a}(\kappa_u+\kappa_d-\kappa_s),
        \nonumber\\
        &
       { \delta M_{\eta a}=\sqrt{\frac{2}{3}}\frac{F_\pi}{F_a}\frac{m_u m_d m_s}{(m_u+m_d)(m_u m_d +m_d m_s+m_s m_u)}m_\pizero^2} \ .
\end{align}

\subsection{$ \Delta S=1$ transitions}
Based on Cirigliano et al \cite{Cirigliano:2011ny} (see also references therein), at the low energy the two operators responsible for $\Delta S=1$ transitions are\footnote{{As mentioned in Sec. \ref{Sec.ChiralGN}, we are working on the phase convention $K_L\simeq \frac{K^0+\bar K^0}{\sqrt 2}$ and $K_S\simeq \frac{K^0-\bar K^0}{\sqrt 2}$.}}
\begin{align}
{\cal L}_{\Delta S=1}=G_8 F_{\pi}^4{\rm Tr}[\lambda_{sd} D^\mu \Sigma^\dag D_\mu \Sigma]
+G_{27}F_{\pi}^4\left(L_{\mu 23} L^\mu_{ 11}+\frac{2}{3}L_{\mu 21} L^\mu_{ 13}\right) +h.c.\,,
\end{align}
where \begin{align}
&L_\mu \equiv i \Sigma^\dag D_\mu \Sigma,\quad
 \lambda_{sd}\equiv\frac{\lambda_6-i\lambda_7}{\sqrt 2}=
\begin{pmatrix}
0&0&0\\
0&0&1\\
0&0&0 
\end{pmatrix}. 
\end{align}
The coefficients $G_{8,27}$ can be determined by the measurement of Kaon decays to pions (see Eq.~\eqref{eq:G8G27} for their value). In order to study the width of $K\to \pi a$ arising from ALP-meson mixing, we need to obtain the trilinear interactions of $K$-$\pi$-$\pi^0/\eta$. 
Below, we expand the two relevant terms in the chiral Lagrangian.

\subsection*{G8 term}

\begin{align}
\frac{i F_\pi G_8}{3 \sqrt{2}}
&\Bigg\{\sqrt{2} \pi ^-
   K^+ \left(2 \sqrt{3} \eta \left(m_\eta^2-m_{K^+}^2\right)+3 {\pi^0} \left(m_{\pi^+}^2-m_{\pi^0}^2\right)\right)
   \nonumber\\
   &+6 K^- K^+ K^0
   \left(m_{K^0}^2-m_{K^+}^2\right)
   \nonumber\\
   &+K^0 \left(-2 \sqrt{3}
   {\pi^0}\eta  \left(m_\eta^2-m_{K^0}^2\right)+3 {\pi^0} {\pi^0} \left(m_{\pi^0}^2-m_{K^0}^2\right)\right)
      \nonumber\\
   &+6K^0 \pi ^- \pi ^+  \left(m_{\pi^+}^2-m_{K^0}^2\right)
   \Bigg\}+h.c.
   \label{Eq:G8}
      \end{align}

\subsection*{G27 term}
\begin{align}
\frac{i F_\pi G_{27}}{9 \sqrt{2}}
&\Bigg\{
\sqrt{6} 
   K^+ \pi ^- \eta \left(4 m_\eta^2-7
   m_{K^+}^2+3 m_{\pi^+}^2\right)
   \nonumber\\
&+3\sqrt{2} 
   K^+ \pi ^- {\pi^0}   \left(-5
   m_{K^+}^2-2 m_{\pi^0}^2+7 m_{\pi^+}^2\right)
   \nonumber\\
  & +12 K^- K^+ K^0 \left( m_{K^0}^2-
   m_{K^+}^2\right)
   \nonumber\\&
   + K^0 \eta\Big(-6 \eta
   \left(m_{K^0}^2-m_\eta^2\right)-3\sqrt{3}  {\pi^0} \left(-2 m_\eta^2
   +m_{K^0}^2+m_{\pi^0}^2\right)\Big)
   \nonumber\\&
   + K^0 \Big(9 {\pi^0}{\pi^0}
   \left(m_{K^0}^2-m_{\pi^0}^2\right)-12 \pi ^- \pi ^+
   \left(m_{K^0}^2-m_{\pi^+}^2\right)\Big)
   \Bigg\}
 +h.c.\label{Eq:G27}
\end{align}

\subsection{$K-\pi-a$ interactions}
From the obtained three-point SM meson interactions reported in the previous section, {we can obtain the three-point ALP-meson-meson interaction via the leading order rotation}    
\begin{align}
\pi^0 &\to \pi^0_{\rm phy}+\theta_{\pi a} a_{\rm phy}\,,\\
\eta&\to \eta_{\rm phy}+\theta_{\eta a} a_{\rm phy}\,.
\end{align}

{In the chiral Lagrangian, the masses come} from derivative-squared terms. If we expand the pion/eta fields to physical pion/eta and axion fields, the axion mass dependence appears due to derivative acting on the axion.  
For example, 
\begin{align}
&(c_1 m_K^2+c_2 m_\pizero^2+c_3 m_\pip^2)K^+ \pi^-  \pizero
&\to& ~~~
\theta_{\pi a}(c_1  m_K^2+c_2 m_a^2+c_3 m_\pip^2)  
 K^+ \pi^- a_{\rm phy}\,, 
\\
&(c_4  m_K^2+c_5 m_\pizero^2)K^0 \pi^0  \pi^0
&\to& ~~~
\theta_{\pi a}(2c_4 m_K^2+c_5 m_a^2+c_5 m_\pizero^2)  
 K^0 \pi^0_{\rm phy} a_{\rm phy}\,. 
\end{align}

The $G_8$ and $G_{27}$ terms in (\ref{Eq:G8}), (\ref{Eq:G27}) with the ALP-meson mixing will lead to the interactions relevant to the $K\to \pi a$ decays (for simplicity, we omit the subscript ``phy''),  
%
\begin{align}
{\cal L}_{\Delta S=1}\to& 
{- iG_8 F_\pi K^+  \pim a \left({-[m_\pip^2-m_{a}^2]\theta_{\pi a}
+ \frac{2}{\sqrt{3}}[m_{K^+}^2-m_a^2]\theta_{\eta a}} \right) 
}\\\nonumber&
{{- iG_8 F_\pi K^0 \pizero a \left(\frac{1}{\sqrt{2}}{[2m_{K^0}^2-m_{\pi^0}^2-m_a^2] }  \theta_{\pi a}
+{\sqrt{\frac{2}{3}}[-m_{K^0}^2+m_a^2]}\theta_{\eta a}
   \right)}
}   
\\\nonumber& 
{- iG_{27}F_\pi K^+ \pi^- a \left(\frac{1}{3}[5m_{K^+}^2+2m_a^2-7m_\pip^2]\theta_{\pi a} 
+{{\frac{1}{3\sqrt{3}}[7m_{K^+}^2-4m_a^2-3m_{\pi^+}^2]}\theta_{\eta a}} \right) \nonumber}
\\\nonumber&
{ -iG_{27}F_\pi K^0 \pizero a \left({-\frac{1}{\sqrt{2}}[2m_{K^0}^2-m_{\pi^0}^2-m_a^2] }  \theta_{\pi a}
{+\frac{1}{\sqrt{6}}[m_{K^0}^2-2m_a^2+m_\pizero^2]\theta_{\eta a}}
\right)  
}+h.c.
\end{align}
%
This leads to the interactions reported in Sec.~\ref{subsec:GG}.


\subsection{Octet Enhancement in   $K^+ \to \pi^+ a$}\label{Sec:AppendixC4}
 \begin{figure*}[t!]
 \begin{center}\label{BRvsBRnaive}
    \includegraphics[width=0.45\linewidth]{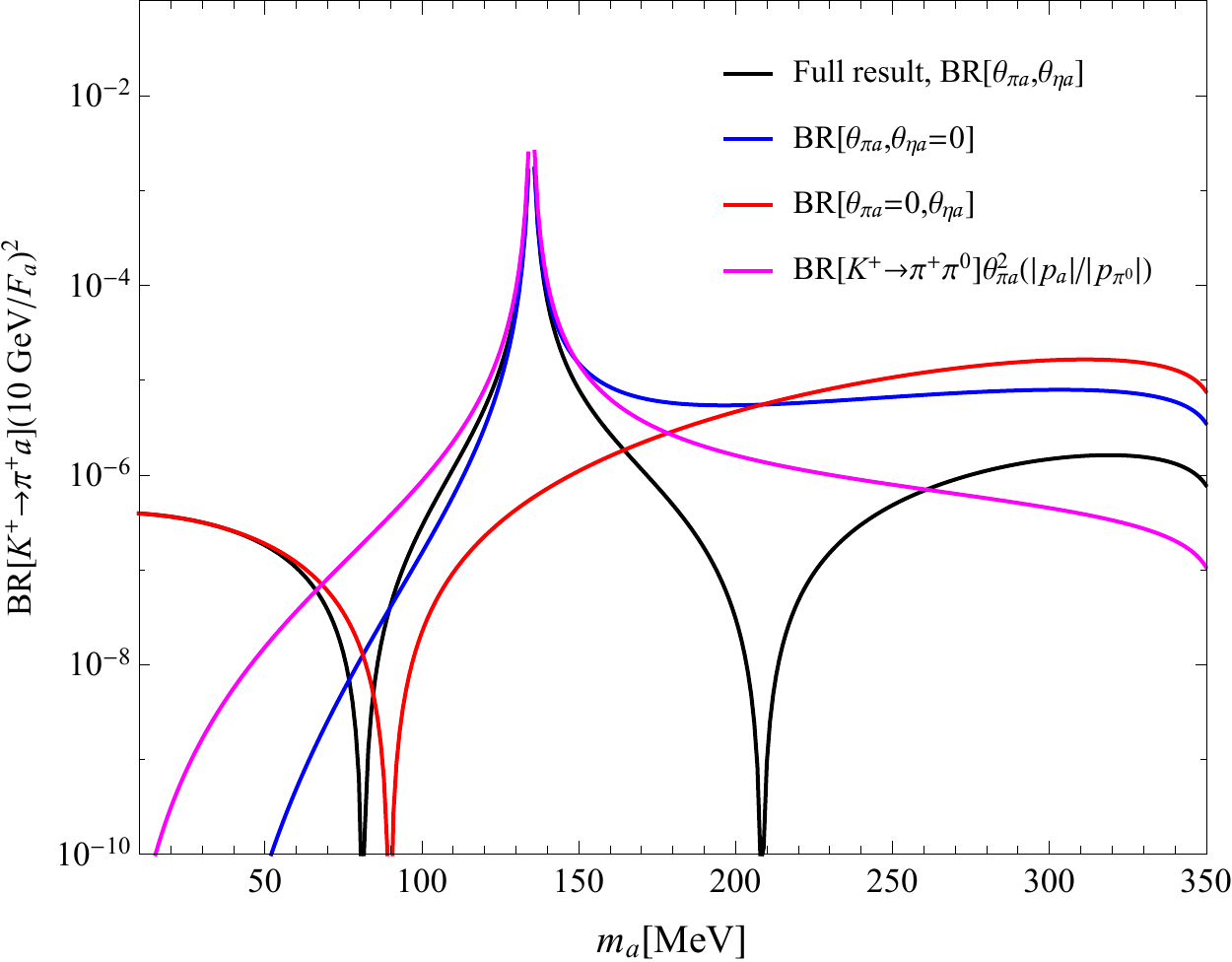}~~~
        \includegraphics[width=0.45\linewidth]{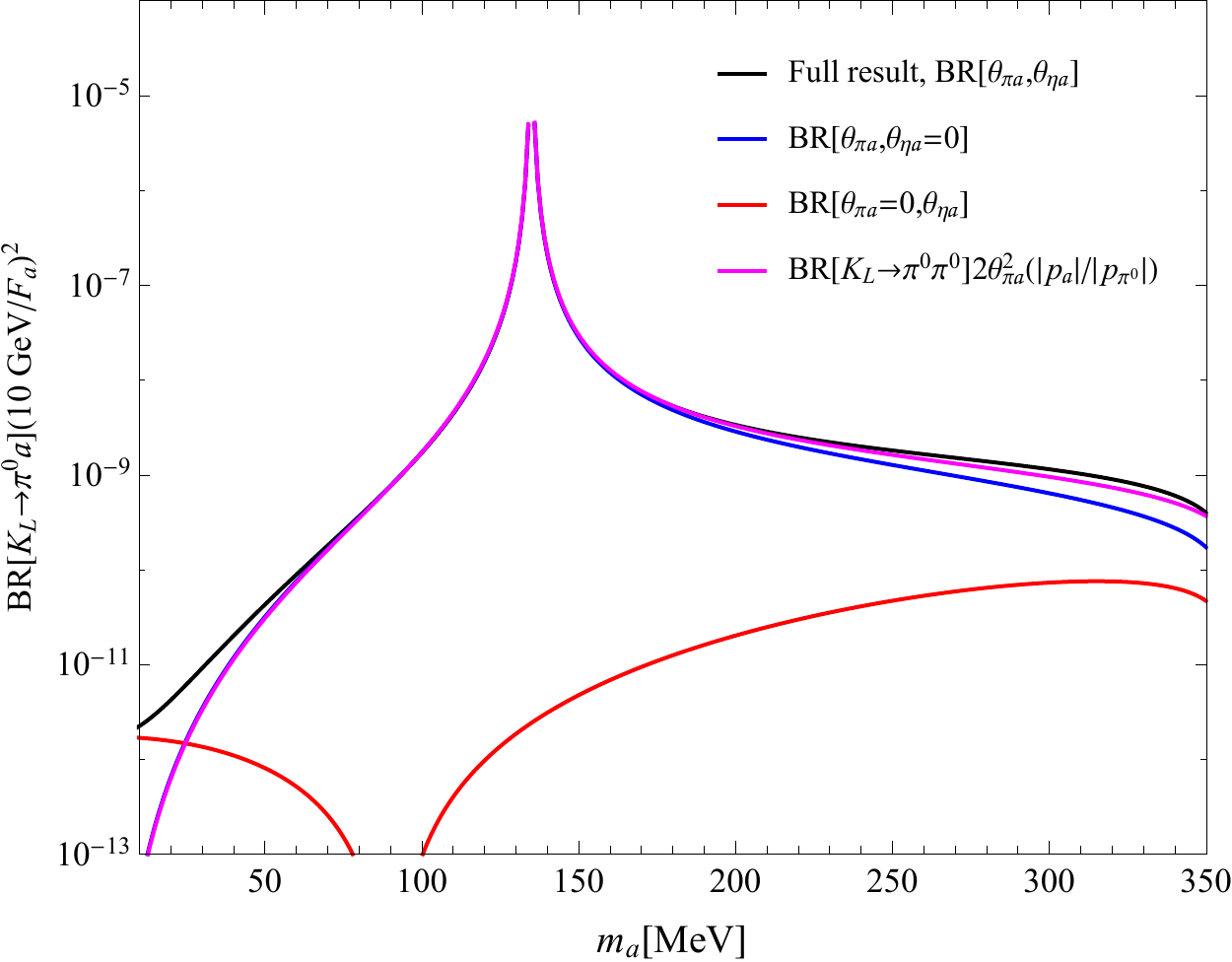}
 \end{center}
 \caption{Left: comparison of our full calculation for $\BR(K^+\to \pi^+ a)$ (in black) to other estimates. The naive scaling of the SM $\BR(K^+\to \pi^+ \pi^0)$ is shown in pink (see (\ref{eq:naiveKplus})). The result that keeps into account only the ALP-pion mixing is shown in blue, the one with only the ALP-eta mixing is shown in red. 
The analogous comparison of the  $\BR(K_L \to \pi^0 a)$ calculations is reported on the right panel. 
}
\end{figure*}
The naive estimate for the $\BR(K^+\to \pi^+ a)$ is often obtained by simply utilizing the ALP-pion mixing. This would lead to
\begin{align}
\BR(K^+\to \pi^+ a)_{\rm{naive}}\simeq \BR(K^+\to \pi^+ \pi^0)\theta_{\pi a}^2 \frac{|\vec{p_a}|}{|\vec{p_\pizero}|}\,,
\label{eq:naiveKplus}
\end{align}
where $\BR(K^+\to \pi^+ \pi^0)\simeq 21\%$, {and $|\vec{p}_a|$ ($|\vec{p}_\pizero|$) is the absolute value of the momentum of the ALP (pion).} 
This, however, only captures a small part of the overall NP effect. In the SM, the $K^+\to \pi^+ \pi^0$ transition is dominated by the $G_{27}$ term while the $G_8$ term is isospin breaking and suppressed by the pion mass splitting (see Eqs. (\ref{Eq:G8}) and (\ref{Eq:G27})). However, if the $\pi^0$ is replaced by the ALP via $\theta_{\pi a}$, there is no such suppression of the $G_8$ term. 
Moreover,  the $G_8$ contribution coming from the $\eta$-ALP mixing also leads to an important contribution \cite{Bardeen:1986yb, Alves:2017avw}. 

In the left panel of Fig.~\ref{BRvsBRnaive}, we numerically compare the naive estimate from Eq.~\eqref{eq:naiveKplus} (pink line) to our full result (black line).
The result obtained keeping only the mixing $\theta_{\pi a}$ ($\theta_{\eta a}$) is also shown in blue (red). 
A large difference between the pink and the black lines is particularly observable at low values of $m_a$. This is due to the fact that the $\eta$ contribution comes in part from ALP-$\eta$ mass mixing that, contrary to kinetic mixing, does not go to zero for $m_a\to 0$.
Furthermore, generically, the two new $G_8$ contributions from $\theta_{\pi a}$ and $\theta_{\eta a}$ have a similar size and can lead to large cancellations {depending on the value of the ALP mass.} However, the value of $m_a$ at which this cancellation happens strongly depend on the various parameters. 
In Fig. \ref{Fig:BoundGG} of the main text, we estimated the uncertainty on the $K^+$ experimental bounds coming from the uncertainty on the quark masses. In addition, there are additional sizable uncertainties coming from the uncertainty on the $\eta$-$\eta'$ mixing angle. 

On the other hand, the prediction on $\BR(K_L\to \pi^0 a)$ is rather stable against these uncertainties because the dominant contribution comes from  the $G_8$ contribution from the ALP-pion mixing, and even the naive formula analogous to Eq.~\eqref{eq:naiveKplus} can typically capture it (see the right panel of Fig. \ref{BRvsBRnaive}). 
Therefore, the prediction on $\BR(K^+\to \pi^+ a)$ has a large uncertainty while the prediction of $\BR(K_L \to \pi^0 a)$ is more theoretically stable.


\subsection{Possible UV contributions to $K \to \pi a$}\label{Appendix E}

Our analysis in Sec. \ref{subsec:GG} for the ALP coupled to gluons assumed that the effective Lagrangian in (\ref{eq:eftGG}) is given at the low energy scale $\mu\sim m_a$. Starting with this Lagrangian, we have shown that the BR$(K_L\to \pi^0 a)$ will be quite suppressed, if compared to BR$(K^+\to \pi^+ a)$ because the former is CP violating (see the $\epsilon_K$ suppression in Eq. (\ref{eq:KLDecayGG})). 

However, {UV completions of this effective Lagrangian generically lead to an enhancement of the branching ratio of the $K_L$ mode at the two-loop order with direct CP violation,} which we schematically described in the following.

The $a\tilde GG$ coupling will also induce a coupling of the ALP with quarks. This is given by
\beq
g_{aqq}^{\rm{eff}} ~a\bar q\gamma_5 q,~~~~g_{aqq}^{\rm{eff}}=-\frac{\alpha_s}{\pi}m_q g_{ag}\left(\log\frac{\mu^2}{m_q^2}-\frac{11}{3}+g(\tau_q)\right)\,,
\eeq
where $\tau_q=4m_q^2/m_a^2$ and $g(\tau)$ is a loop function that can be found e.g. in \cite{Bauer:2017ris}. Thanks to these induced coupling, the ALP will be produced in $K_L\to\pi^0 a$ through penguin diagrams with the ALP radiated from the quark loop. The corresponding partial width {can be estimated as~\cite{Wise:1980ux}}

\beq
\Gamma(K_L\to\pi^0 a){\approx\frac{G_F^2}{(2\pi)^4}}f_L^2m_{K_L}^3\sum_{\alpha\in c,t}{\rm{Im}}\left((V_{\alpha d}V_{\alpha s}^*)^2\right)(g_{a\alpha\alpha}^{\rm{eff}})^2m_\alpha^2\log\frac{m_W^2}{m_\alpha^2},
\eeq
where $f_L$ is the form factor in the $K_L$ decay, and $V_{\alpha d},~V_{\alpha s}$ are CKM factors. This decay is induced, for example, by the dimension five operator $\partial_\mu a (\bar{s}_L \gamma_5\gamma^\mu d_L)$  as in the SU(2) coupled ALP case of Sec. \ref{Sec:SU(2)}. In this case, we will not have a $\epsilon_K$ suppression in the decay. Thus, this two-loop contribution can potentially significantly enhance the $\BR(K_L\to \pi^0 a)$ (see also \cite{Choi:2017gpf}). 

Similarly, also the rate of the charged mode $K^+\to\pi^+a$ can be enhanced by these UV contributions: 
\beq
\Gamma(K^+\to\pi^+ a){\approx\frac{G_F^2}{(2\pi)^4}}f_+^2m_{K^+}^3\sum_{\alpha\in c,t}
\left|V_{\alpha d}V_{\alpha s}\right|^2(g_{a\alpha\alpha}^{\rm{eff}})^2m_\alpha^2\log\frac{m_W^2}{m_\alpha^2},\eeq
{The enhancement is, however, not as sizable as the one in the neutral mode, because of the absence of the $\epsilon_K$ suppression in the IR contribution.}

\bibliographystyle{utphys}
\bibliography{KOTObib}

\end{document}